\documentclass[compsoc, conference, a4paper, 10pt, times]{IEEEtran}
\IEEEoverridecommandlockouts

\usepackage{cite}
\usepackage{amsthm,amsmath,amsfonts}
\usepackage{mathtools}
\usepackage[boxruled,titlenotnumbered]{algorithm2e}
\usepackage{algorithmicx}
\usepackage{algpseudocode}
\usepackage{graphicx}
\usepackage{textcomp}
\usepackage{xcolor}
\usepackage[binary-units=true]{siunitx}
\usepackage{refcount}
\usepackage{xspace}
\usepackage{url}

\usepackage{breakurl}
\usepackage{boxedminipage}
\usepackage{subcaption} %

\definecolor{myblue}{RGB}{70,130,180}
\definecolor{mydeepblue}{RGB}{65,105,225}
\definecolor{myviolet}{RGB}{97,0,138}
\definecolor{myburgundy}{RGB}{110,10,30}
\definecolor{mygreen}{RGB}{0,105,148}
\definecolor{mygrey}{RGB}{180, 180, 200}
\definecolor{idealfun}{RGB}{165,42,42}
\definecolor{check}{RGB}{11,141,10}
\definecolor{cross}{RGB}{223,68,52}
\definecolor{grayhighlight}{RGB}{250,250,227}
\definecolor{plum}{RGB}{209,154,212}

\usepackage[ff,mm,sets,adversary,advantage,asymptotics,probability,keys,primitives,operators,lambda,logic,notions]{cryptocode}

\colorlet{party}{myburgundy}
\colorlet{protstring}{myviolet}
\colorlet{comment}{mygrey}

\newcommand{\cmt}[1]{\textcolor{comment}{\text{// #1}}}

\newcommand{\PROT}{\mathit{Prot}}

\newcommand{\RP}{\textcolor{party}{\mathcal{RP}}\xspace}
\newcommand{\IDP}{\textcolor{party}{\mathcal{IDP}}\xspace}
\newcommand{\mixer}{\textcolor{party}{\mathcal{M}}\xspace}
\newcommand{\user}{\textcolor{party}{\mathcal{U}}\xspace}
\newcommand{\UA}{\textcolor{party}{\mathcal{UA}}\xspace}

\newcommand{\onrecv}{\textcolor{mygreen}{\textbf{On receive}}}
\newcommand{\oninit}{\textcolor{mygreen}{\textbf{On initialization}}}
\newcommand{\msg}[1]{\textcolor{protstring}{\text{``#1''}}}

\newcommand{\gatt}{\textcolor{idealfun}{\mathcal{G}_{\text{att}}}\xspace}
\newcommand{\randseed}{n}

\newcommand{\program}{\mathsf{prog}}
\newcommand{\inputtee}{\mathsf{inp}}
\newcommand{\outputtee}{\mathsf{outp}}
\newcommand{\sigschemetee}{\sum_{\mathsf{TEE}}}
\newcommand{\sigtee}{\sigma_{\mathsf{TEE}}}
\newcommand{\sktee}{\mathsf{sk_{TEE}}}
\newcommand{\pktee}{\mathsf{pk_{TEE}}}
\newcommand{\pktls}{\mathsf{pk_{TLS}}}

\newcommand{\eid}{\mathsf{eid}}

\usepackage{pgfplots}
\usetikzlibrary{patterns}
\usepackage{tabularx}
\usepackage{diagbox}
\usepackage{booktabs}
\usepackage{threeparttable}
\usepackage{xcolor, colortbl}
\usepackage{tikz}
\usepackage{enumitem}
\usepackage[T1]{fontenc}
\usepackage{bbding}
\usepackage{pifont}
\usepackage{amssymb}
\newcommand{\cmark}{\textcolor{check}{\ding{51}}}%
\newcommand{\xmark}{\textcolor{cross}{\ding{55}}}%

\usepackage{dtrt}

\newcommand{\req}[1]{\emph{Q#1}}

\definecolor{camel}{rgb}{0.76, 0.6, 0.42}

\newcommand{\boldhead}[1]{\parhead{#1}}

\newcommand{\name}{\textsf{MISO}\xspace}
\newcommand{\mofn}[2]{${#1}\text{-of-}{#2}$}
\newcommand{\extendedprotocol}{Multi-IdP SSO}

\newcommand{\redirecturiB}{\texttt{uri\_mixer}}
\newcommand{\redirecturiRP}{\texttt{uri\_RP}}
\newcommand{\clientidB}{\texttt{$\mathtt{cid_{mixer}}$}}
\newcommand{\clientidRP}{\texttt{$\mathtt{cid_{RP}}$}}
\newcommand{\clientsecB}{\texttt{$\mathtt{csec_{mixer}}$}}
\newcommand{\clientsecRP}{\texttt{$\mathtt{csec_{RP}}$}}
\newcommand{\localstateB}{\texttt{$\mathtt{state_{mixer}}$}}
\newcommand{\localstateRP}{\texttt{$\mathtt{state_{RP}}$}}
\newcommand{\codeB}{\texttt{$\mathtt{code_{mixer}}$}}
\newcommand{\codeRP}{\texttt{$\mathtt{code_{RP}}$}}
\newcommand{\tokenB}{\texttt{$\mathtt{token_{mixer}}$}}
\newcommand{\tokenRP}{\texttt{$\mathtt{token_{RP}}$}}
\newcommand{\authendB}{\texttt{/auth\_mixer}}
\newcommand{\authendIdP}{\texttt{/auth\_IdP}}
\newcommand{\tokenendB}{\texttt{/token\_mixer}}
\newcommand{\tokenendIdP}{\texttt{/token\_IdP}}
\newcommand{\resourceendB}{\texttt{/res\_mixer}}
\newcommand{\resourceendIdP}{\texttt{/res\_IdP}}
\newcommand{\userid}{\texttt{$\mathtt{uid}$}}
\newcommand{\uid}{\texttt{$\mathtt{UID}$}}
\newcommand{\preuid}{\texttt{$\mathtt{preUID}$}}
\newcommand{\idplist}{\texttt{$\mathtt{idp\_list}$}}
\newcommand{\ktee}{\mathsf{sk^{PRF}_{TEE}}}

\newcommand{\etal}{\emph{et al.}\xspace}

\theoremstyle{definition}
\newtheorem{definition}{Definition}

\newcommand{\mysquare}[1]{
  \hbox{%
    \color{mydeepblue}
    \vrule width2.5ex height1.85ex depth.4ex%
    \hskip-2.5ex%
    \hbox to2.5ex{%
      \hfil%
      \color{white}#1%
      \hfil}%
  }%
}

\newcommand \footnoteONLYtext[1]
{
	\let \mybackup \thefootnote
	\let \thefootnote \relax
	\footnotetext{#1}
	\let \thefootnote \mybackup
	\let \mybackup \imareallyundefinedcommand
}

\begin{document}

\title{\name{}: Legacy-compatible Privacy-preserving Single Sign-on\\ using Trusted Execution Environments}

\iftrue
\author{
\IEEEauthorblockN{
Rongwu Xu\IEEEauthorrefmark{1},
Sen Yang\IEEEauthorrefmark{2},
Fan Zhang\IEEEauthorrefmark{2},
Zhixuan Fang\IEEEauthorrefmark{1}\IEEEauthorrefmark{4}
}

\IEEEauthorblockA{
\IEEEauthorrefmark{1}Institute for Interdisciplinary Information Sciences, Tsinghua University,\\ \IEEEauthorrefmark{2}Department of Computer Science, Yale University, \IEEEauthorrefmark{4}Shanghai Qi Zhi Institute
}

\IEEEauthorblockA{
xrw22@mails.tsinghua.edu.cn, \{sen.yang, f.zhang\}@yale.edu, zfang@mail.tsinghua.edu.cn
}
}

\fi

\maketitle

\begin{abstract}
Single sign-on (SSO) allows users to authenticate to third-party applications through a central identity provider.
Despite their wide adoption, deployed SSO systems suffer from privacy problems such as user tracking by the identity provider. 
While numerous solutions have been proposed by academic papers, none were adopted because they require modifying identity providers, a significant adoption barrier in practice. 
Solutions do get deployed, however, fail to eliminate major privacy issues.

Leveraging Trusted Execution Environments (TEEs), we propose \name{}, the first privacy-preserving SSO system that is completely compatible with existing identity providers (such as Google and Facebook).
This means \name{} can be easily integrated into existing SSO ecosystem today and benefit end users.
\name{} also enables new functionality that standard SSO cannot offer: \name{} allows users to leverage multiple identity providers in a single SSO workflow, potentially in a threshold fashion, to better protect user accounts.
We fully implemented \name{} based on Intel SGX. Our evaluation shows that \name{} can handle high user concurrency with practical performance.
\end{abstract}

\section{Introduction}
\label{sec:intro}

\begin{figure*}
  \centering
  \subfloat[SSO]{\includegraphics[width=0.42\textwidth]{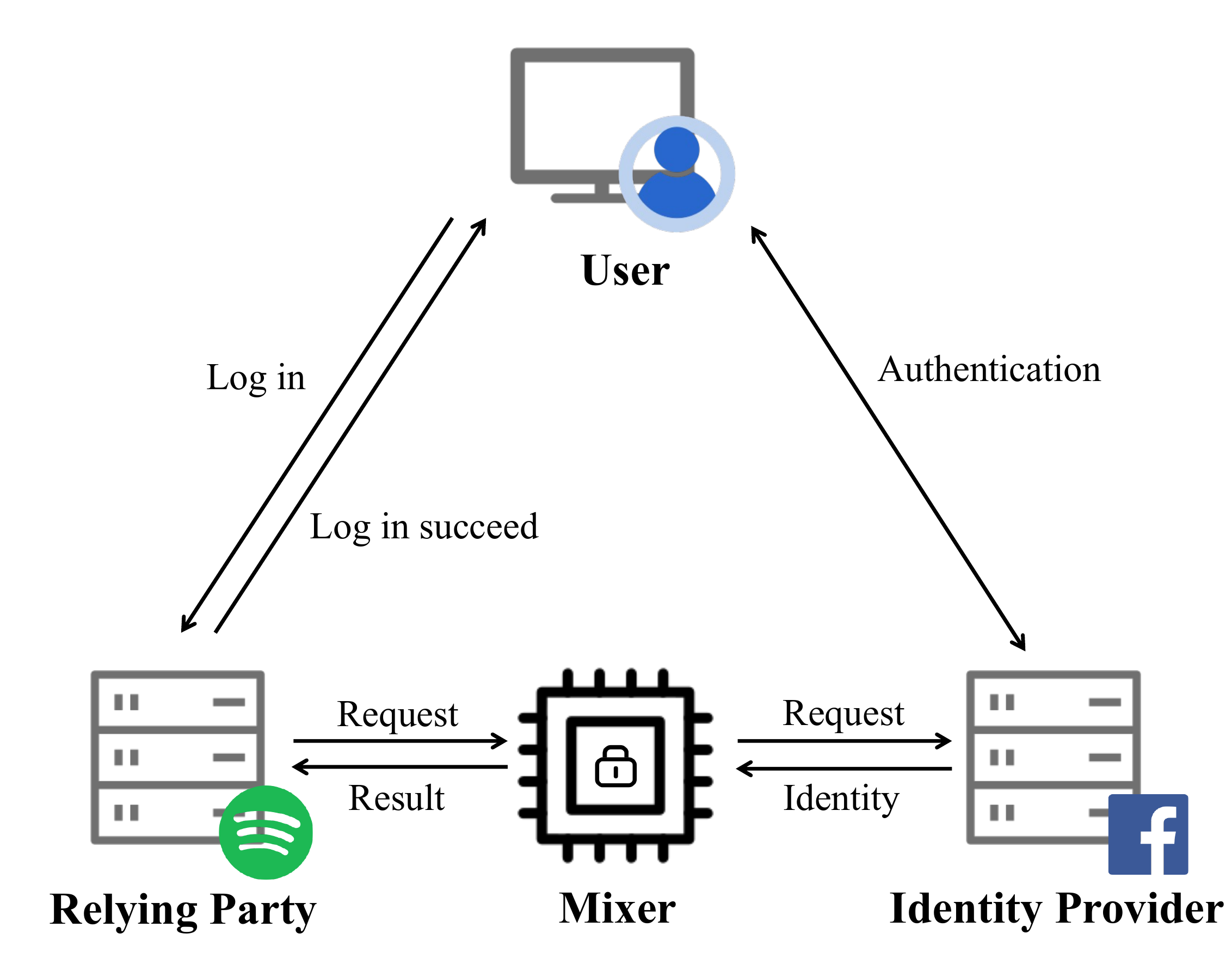}\label{fig:single}}
  \hspace{1cm}
  \subfloat[\extendedprotocol{}]{\includegraphics[width=0.42\textwidth]{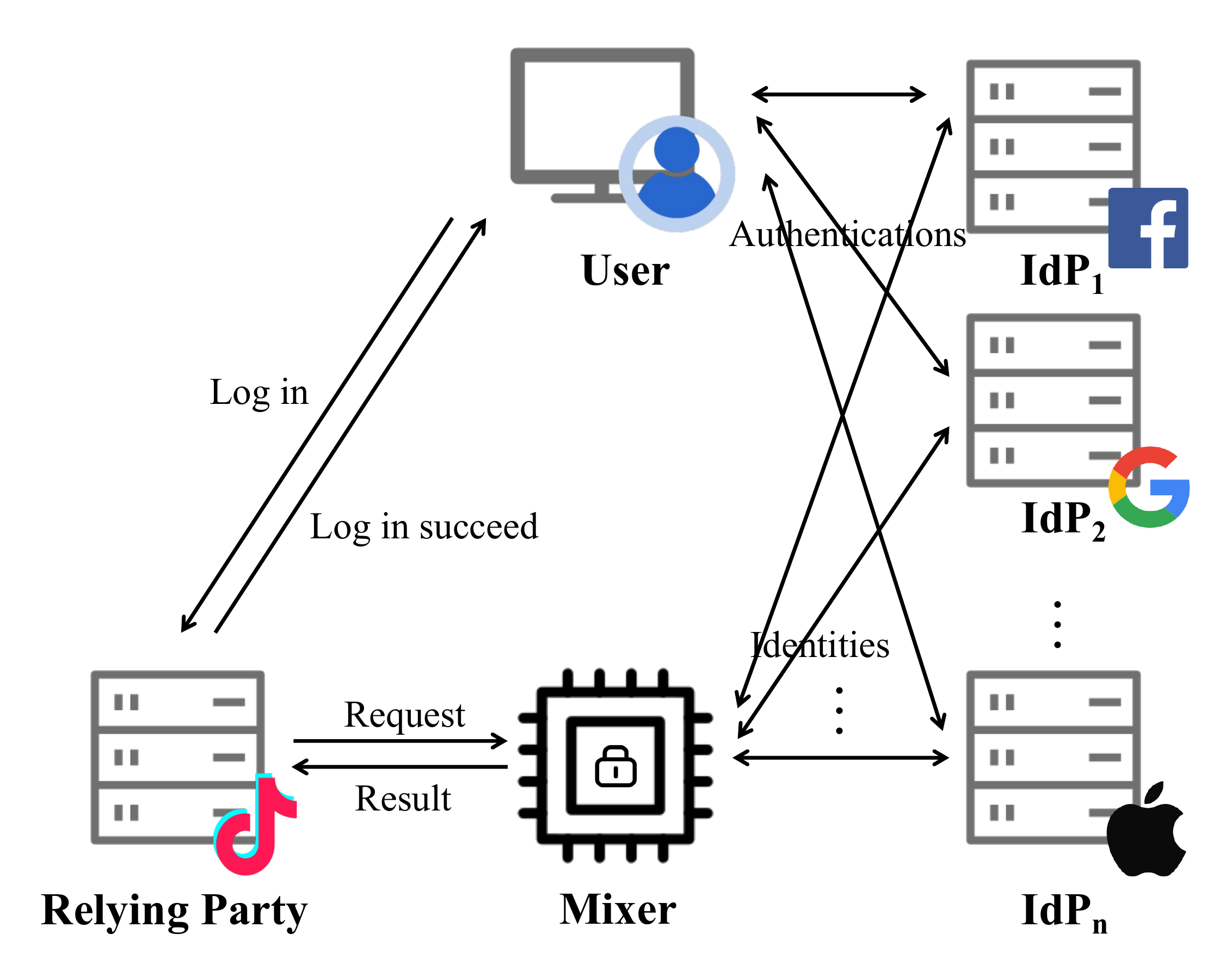}\label{fig:multiple}}
  \caption{Login with \name{} under different settings: (a) a single IdP; (b) multiple IdPs.}
  \label{fig:intro}
\end{figure*}

``The end of passwords'' is nominated as one of the ten breakthrough technologies in 2022 by MIT Technology Review~\cite{mittech}, as its wide adoption may finally put an end to the perennial breaches caused by improper uses of passwords~(see, e.g., \cite{klein1990foiling} for a survey). Among the proposed passwordless authentication techniques (e.g., biometrics, hardware tokens, magic links, etc.), \emph{single sign-on (SSO)}---with which users can authenticate to third-party applications through a central identity provider---stands out with unique advantages: it offers a familiar user experience, it does not require special hardware, and it is already widely supported.
As of 2022, more than 84,000 websites support SSO through Google and more than 80,000 through Facebook~\cite{similartech}. 

At a higher level, SSO is a three-party protocol among a \emph{user}, an \emph{Identity Provider} (referred to as IdP for short), and a third-party application (referred to as a \emph{Relying Party} or RP) to which the user wishes to authenticate. In order for the RP to authenticate the user, it redirects the user to an IdP, who authenticates the user through established means (e.g., password, QR codes, etc.) and informs the RP of the user's identity (typically by providing a token).
One of the most widely adopted SSO protocols today is OAuth 2.0 \cite{oauthweb, hardt2012rfc}, supported by popular IdPs like Facebook, Twitter, Apple, and Microsoft \cite{facebooklogin, loginwithtwitter, signinwithapple, loginwithtmicrosoft}.

While SSO is widely adopted, deployed systems suffer from several well recognized security and privacy problems~\cite{maler2008venn, uruena2010analysis, sun2011makes, wang2011third, gafni2014social, fett2015analyzing, li2020user,morkonda2021empirical}: 

\begin{itemize}[leftmargin=*]
\newcommand{\localtextbulletone}{\textcolor{black}{\raisebox{.2ex}{\rule{.8ex}{.8ex}}}}
\renewcommand{\labelitemi}{\localtextbulletone}
    \item \emph{Account linkage by identity providers (IdPs)}~\cite{sun2011makes,gafni2014social,fett2015analyzing,li2020user}. Positioned at the center of the SSO workflow, identity providers (IdPs) have significant surveillance power. In particular, they can track which websites users visit, contributing to the massive scale of data that big tech firms already possess. 
        
    \item \emph{Account linkage across relying parties (RPs)}~\cite{maler2008venn}\cite[\S 17.3]{oidcfinal}\cite[\S 3.3]{hammann2020privacy}. In existing SSO systems~\cite{googleoauthwithopenid, facebooklogin}, IdPs identify users with the same unique user identifiers across RPs, which allows a group of colluding RPs to track a given user's login activities across websites.
    
    \item \emph{Unnecessary identifier exposure to RPs}\cite{maler2008venn,uruena2010analysis,sun2011makes,wang2011third,morkonda2021empirical}. While existing SSO systems give users some control over what to disclose to the RPs, many reveal unique identifiers (e.g., email) unconditionally. This is neither necessary for many applications nor desirable. For example, 
    Tiktok \cite{tiktok} does not need user emails, but is nonetheless provided with one when a user registers through Google SSO.

    \item \emph{Single-point failures of account security and availability}~\cite{sun2011makes,gafni2014social}. As the name suggests, SSO relies on a single IdP, which implies strong assumptions about the security and availability of the account on that IdP. This can be undesirable for crucial applications (e.g., cryptocurrency wallet~\cite{TorusLabsOpenSourceKeyManageme}), where account compromise can be catastrophic. Service downtime can also render users unable to access third-party applications. A better approach is to allow users to choose multiple IdPs (potentially in a threshold fashion for better availability). Although RPs often offer several IdP choices, users can only select a single IdP to log in, and each IdP option creates a separate account. 
    So far, there is no solution to support multiple IdPs while achieving the aforementioned privacy goals.
\end{itemize}

It is important to note that these issues are well recognized and several solutions have been proposed (e.g., SPRESSO \cite{fett2015spresso}, PRIMA \cite{asghar2018prima}, POIDC \cite{hammann2020privacy}, EL PASSO \cite{zhang2021passo}, and UPPRESSO \cite{guo2021uppresso}). However, to the best of our knowledge, none of these academic solutions has seen the light of real-world adoption. A key reason is that they lack \emph{legacy compatibility}---these solutions require the identity providers to deploy new software, which creates a significant adoption barrier.

Meanwhile, solutions do get deployed only solved part of the problems. For example, OIDC with Pairwise Pseudonymous Identifiers introduced in OpenID Connect 1.0~\cite{sakimura2014openid} can prevent account linkage across RPs, but it does not eliminate other privacy issues.
Similarly, Apple~\cite{signinwithapple, signinwithappleemailrelay} launched ``Sign in with Apple'' in 2019, an SSO service that auto-generates a random forward email address for each RP to hide user email and prevent cross-RP linkage, but Apple (as the IdP) can still track users' login activities.

In this paper, we aim to combine the best of both worlds: {\em can we design an SSO protocol that not only achieves strong privacy guarantees, but also remains fully compatible with existing identity providers?}

\boldhead{Our solution: \name{}.}
We start with addressing the issue of account linkage by IdP and extend the basic protocol to address other privacy issues.
Our key idea is to build a \textbf{\em mixer} to shield users from identity providers.
(\name stands for a \underline{mi}xer of S\underline{SO}.)
\autoref{fig:single} illustrates the high-level idea.
Suppose the user $\user$ wants to log in to a relying party $\RP$ through a specific identity provider $\IDP$.
Instead of having $\RP$ directly interact with $\IDP$, \name introduces a {\em mixer} $\mixer$, which first authenticates the user through $\IDP$, then informs $\RP$ of the user's identity, finishing the SSO workflow.

From $\IDP$'s point of view, it only observes that the user is logging in to the mixer---the fact that the user is actually logging in to $\RP$ is hidden from $\IDP$. The mixer can further obfuscate the information obtained from $\IDP$ before presenting to $\RP$, to prevent account linkage across RPs, as well as to allow users to apply arbitrary data redaction, e.g., removing emails returned by Sign in with Google.

\boldhead{Challenges and solutions.}
Realizing \name{} faces two challenges.
First, \name{} introduces a new entity outside the specification of SSO, which goes against our goal of remaining completely backward compatible with deployed SSO systems. How can we retrofit the mixer $\mixer$ to the existing SSO architecture in a \emph{transparent} way? Second, if the mixer $\mixer$ is adversarial, it could not only track users, but also mount impersonation attacks, i.e., it can log in to $\RP$ as any user. How do we realize the mixer $\mixer$ in a \emph{trustworthy} fashion? 

To tackle the first challenge, we run two {\em nested} SSO workflows together. The mixer $\mixer$ acts as an identity provider when interacting with $\RP$, while as a relying party when interacting with $\IDP$. This way, the mixer $\mixer$ is realized with two roles that are both compatible with existing SSO infrastructures.

To realize a trustworthy mixer $\mixer$, several options are possible. For example, one could realize the $\mixer$ with multi-party computation (MPC) so that as long as some fractional of MPC nodes are honest, the adversary (who are limited by the number of MPC nodes he can compromise) cannot track or impersonate users. 
While theoretically feasible, even state-of-the-art MPC protocols incur orders of magnitude overhead. 
To achieve practical performance, we realize $\mixer$ with a \emph{Trusted Execution Environment (TEE)}.
To establish secure channels, TEE generates a TLS private key and binds the TEE code to the corresponding public key through remote attestations~\cite{johnson2016intel}. \name{} operates in a trust-upon-first-use model---users (and RPs) verify the attestation the first time they use \name{}, after which they can put the verified public key (and the certificate) in their local keychain.

\boldhead{Supporting multiple identity providers.}
As shown in~\autoref{fig:multiple}, \name{} can be naturally generalized to support multiple IdPs in what we call \extendedprotocol{}.
This enables a user to protect her third-party application accounts with multiple IdPs (e.g., $2$ out of $3$ IdPs) to hedge against account compromises or service downtime. On the one hand, even when some IdPs accounts are compromised, the attacker still can not impersonate her to log in to the RP. On the other hand, when some IdPs are unavailable (e.g., due to service downtime), the user can still access her account.

\boldhead{Implementation and evaluation.} To demonstrate the practicality of \name{}, we implement a prototype of \name{} with the Intel Software Guard Extensions (SGX), a widely available TEE. Note that \name{} is not specific to Intel SGX in any way and can be realized with any TEEs. 
We deploy the prototype of \name{} on a server with 30GB memory and Intel SGX hardware. We evaluate its performance under concurrent user login requests. Under $200$ users/second concurrent login requests, it takes $454$ms to finish the standard SSO login on average, and $773$ms for the \mofn{2}{3} \extendedprotocol{}.
We also compare the performance with a baseline (insecure) implementation without SGX on the same hardware. 
Our results suggest that average latency of the standard \name{} is $2.19 \times$ over the basic SSO, and the overhead introduced by SGX is no more than $15\%$. 
The evaluation suggests that our prototype achieves practical performance.

\boldhead{Contributions and road map.}
In summary, our main technical contributions are:
\begin{enumerate}
    \item \emph{\name{}}. We are the first to design a legacy-compatible privacy-preserving SSO system. Our system guarantees account unlinkability w.r.t. both the IdP and RPs, and allows users to selectively disclose user identity, while remaining completely compatible with existing SSO identity providers.
    We present the design of \name{} and the SSO workflow in \autoref{sec:design}, along with a full protocol specification.
    \item \emph{\extendedprotocol{}}. \name{} enables the user to sign-on with multiple IdPs in a single SSO workflow. This enables users to leverage multiple IdPs to hedge against single-point failures of account security and availability. This feature can also be used to create decentralized identifiers (DIDs)~\cite{maram2021candid}  from multiple providers.
    We specify the \extendedprotocol{} workflow in \autoref{sec:extended}.
    \item \emph{Prototype implementation and evaluation}. We implement a prototype of \name{} and evaluate its performance with concurrent user requests. We also compare the prototype with an (insecure) baseline without TEE. Evaluation of the prototype suggests that \name{} achieves practical performance, with a latency about $2 \times$ over the basic SSO. 
    Implementation and evaluation results are given in \autoref{sec:implement}.
\end{enumerate}

In what follows, we provide backgrounds on OAuth 2.0 and TEE in \autoref{sec:background}, adversary model and design goals in \autoref{sec:goals}.
We present security analysis in \autoref{sec:analysis}.
We discuss extended topics in \autoref{sec:discussion}.
Finally, we summarize and review related works in \autoref{sec:related} and conclude our work in \autoref{sec:conclusion}.

\section{Background}
\label{sec:background}

In this section, we provide the background on the basic properties of SSO authentication, the OAuth 2.0 framework, and TEEs.

\subsection{Properties of SSO}
\label{sec:SSO properties}

The key properties for a common SSO solution are:
\begin{itemize}
    \item \emph{User identification at the RP.} The result of an instance of SSO is that the RP can get a unique identifier of the user.
    \item \emph{User authentication only at the IdP.} In each SSO login instance, the user only authenticates with an IdP once. In particular, the user do not need to authenticate for each RP, or to create and keep any credentials or keys.
    \item \emph{User identity attributes retrieval.} The RP is able to obtain the user's identity attributes confirmed and issued by the IdP.
\end{itemize}

\subsection{OAuth 2.0}
\label{sec:oauth 2.0}

\begin{figure}
  \centering
  \includegraphics[width=\linewidth]{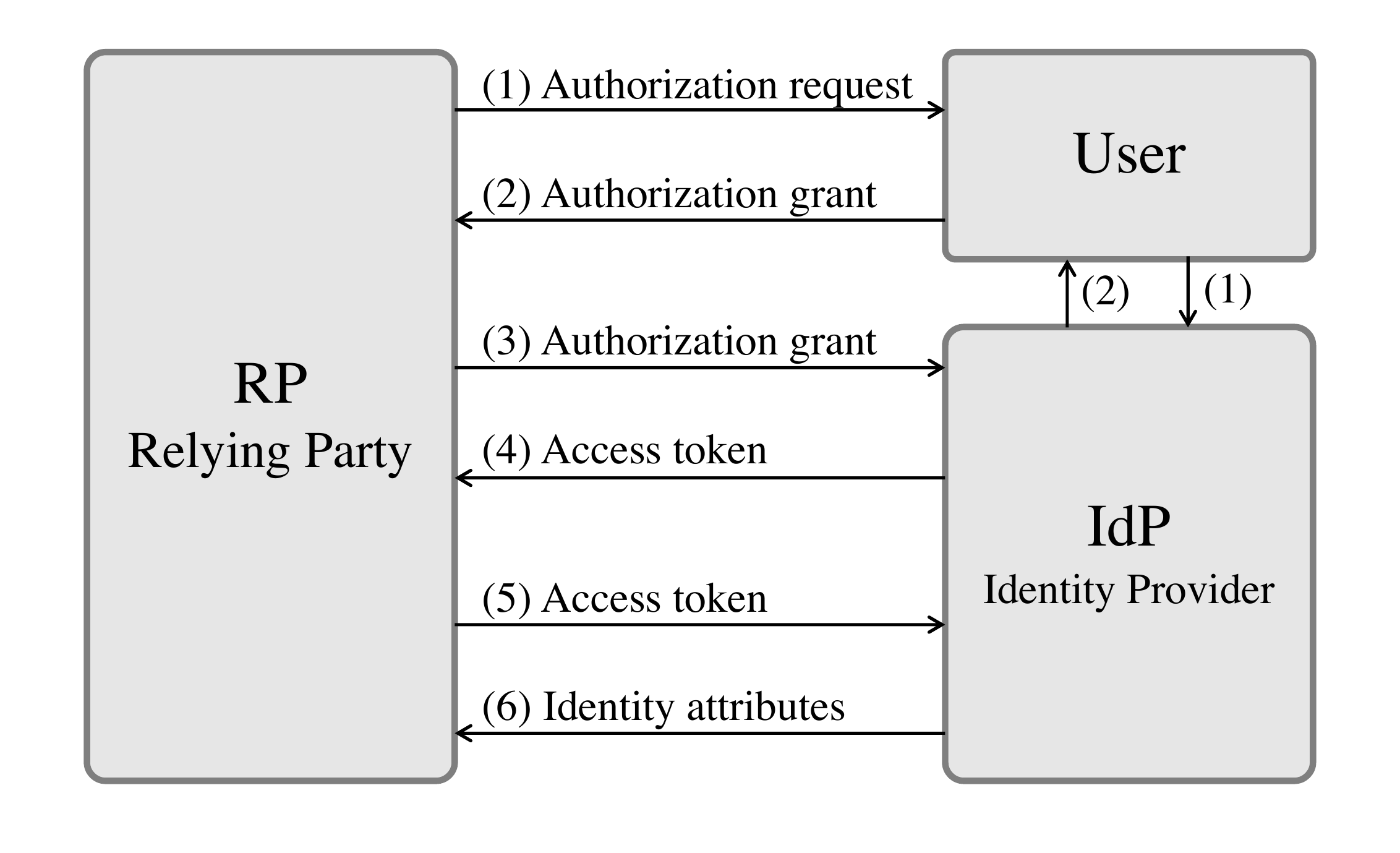}
  \caption{The OAuth 2.0 abstract workflow.}
  \label{fig:oauth}
\end{figure} 

Multiple SSO standards and protocols have been deployed over the years, such as SAML \cite{hughes2005security}, OpenID \cite{recordon2006openid}, OAuth 2.0 \cite{oauthweb, hardt2012rfc} and OpenID Connect (OIDC) \cite{sakimura2014openid, oidcfinal}.
Published in 2012, the OAuth 2.0 has become one of the most prominent SSO protocol frameworks.
OAuth 2.0 allows the RP to retrieve the user's protected resources with restrictions from the IdP, and it can be used for SSO if the IdP returns the user's identity attributes to convince the RP of her identity. 
OAuth 2.0 is also a basis for a widely adopted SSO protocol, OIDC, which adds a simple identity layer over OAuth 2.0.
In most circumstances, an IdP that supports OIDC is also compatible with OAuth 2.0, e.g., Google's ``Sign in with Google'' solution \cite{googleoauthwithopenid}.

As shown in ~\autoref{fig:oauth}, OAuth 2.0 is a three-party protocol involving RP, IdP, and the user. 
In short, RP initiates the workflow and lets the user authenticate at IdP. Then, RP obtains the grant (code) and token from IdP. Finally, RP obtains the user's identity attributes with the token. 
For page limitations, we defer the details to ~\autoref{sec:oauthbackground}. Readers are encouraged to consult since our design utilizes the OAuth 2.0 protocol as its building block. Though the paper itself is self-contained.

\subsection{TEE and Intel SGX}

A trusted execution environment (TEE) offers a secure execution environment that guarantees confidentiality and integrity of the code and data loaded and executed inside it. TEE can be used to run secure applications inside on an untrusted environment (e.g., a host).

Intel Software Guard Extensions (SGX) \cite{intel2015intel,costan2016intel,anati2013innovative,mckeen2013innovative,hoekstra2013using} is an extension of the x86 instruction set architecture, which offers hardware-based memory encryption that isolates specific user-level application code and data in memory. SGX allows applications to create isolated memory regions for code and data (called \emph{enclaves}) to protect them from potential malicious privileged software \cite{wojtczuk2009attacking}, operating system, hypervisor code, or even some hardware attacks \cite{halderman2009lest} on the same host. Code and data inside the enclave are protected in a space of memory pages called enclave page caches (EPC) and encrypted by the memory encryption engine (MEE) to prevent access outside the enclave.
The main protective technologies of SGX include: runtime isolation (enclave), \texttt{ecall/ocall} interfaces, attestation, and sealing, readers may refer to \cite{costan2016intel} for a more in-depth exposition.

\section{Adversarial Model and Design Goals}
\label{sec:goals}

In this section, we begin with the adversarial model and assumptions. Then we state the privacy goals along with other desired properties that \name{} enjoys. 

\subsection{Adversarial Model}
\label{subsec:adversarial-model}

According to \autoref{fig:intro}, there are four parties in \name{}: the IdP, the RP, the user, and the mixer. Since our system is web-based, the user uses a user-agent that can perform actions on her behalf.
We consider the following adversarial model:

\boldhead{Honest-but-curious IdP.} The IdP is honest, which means it is implemented correctly and follows the standard OAuth 2.0 protocol.
But a curious IdP is interested in collecting the user information throughout the execution of the protocol, e.g., tracking the user's login activities among RPs.

\boldhead{Malicious RPs.} The RPs may intentionally deviate from the protocol to investigate interactions between other parties.
The adversary can control a group of RPs to break the privacy guarantees.
We only consider that the group of malicious RPs can only make use of information leaked from the protocol, e.g., they can not make use of users' network connection information.

\boldhead{Users.} Users can be honest or malicious. Malicious users can collude with the RPs. For instance, malicious users and RPs may work together to impersonate victim users in order to log in to other honest RPs.
We assume an honest user deploys a secure user-agent, commonly a modern web browser, which is supposed to safeguard its user's local credentials, e.g., her password on the IdP.
Also, an honest user never allows distinctive identity attributes (personally identifiable information \cite{w3cverifiable}) that can be easily correlated to her, e.g., telephone number, or email address, to be shared. So that unlinkability is possible across the RPs. 

\boldhead{Malicious mixer host.} The mixer is implemented in a TEE. The adversary could compromise the host of the mixer by exploiting software vulnerabilities, and attempt to steal system secrets.
We use Intel SGX in particular for implementation, and we assume the enclave achieves integrity and confidentiality guarantees as defined in \cite{costan2016intel, intelsgxoverview}. 

\boldhead{Correct OAuth 2.0 implementation.} 
Care must be taken to implement and configure OAuth 2.0 correctly. We refer readers to \cite{sun2012devil,fett2016comprehensive,mainka2017sok} for a survey of OAuth 2.0 implementation pitfalls.
\name{} is built on top of up-to-date OAuth 2.0 implementation and we assume the implementation is free of known OAuth 2.0 protocol-level vulnerabilities. 

Denial-of-Service (DoS) attacks are out of scope.

\subsection{Design Goals}
\label{subsec:goals}

\boldhead{Privacy goals.}
We aim to achieve the following privacy goals (formal privacy definitions will be given in~\autoref{subsec:analysis}):
\begin{itemize}
    \item \emph{IdP Unlinkability} (\autoref{def:idpunlink}). Informally, given two RPs $RP_1$ and $RP_2$ and a user who logs in to one of them through an IdP, IdP unlinkability requires that the IdP cannot decide which RP the user logged in to. The definition generalizes the multiple RPs naturally.
    \item \emph{RP Unlinkability} (\autoref{def:rpunlink}). This definition captures the requirement that a group of RPs cannot trace user login activities. More precisely, given two users $U_1$ and $U_2$, when one of them logs in at a given RP, that RP cannot decide which user logged in, even through colluding with other RPs to which $U_1$ and $U_2$ have previously logged in. Note that in existing SSO protocols, RPs learn users' unique identifiers from the IdP, and thus can break this property. 
    \item \emph{Collusive-IdP-RP Unlinkability} (\autoref{def:idprpunlink}). IdP unlinkability and RP unlinkability should hold even when the IdP colludes with  RPs and reveals users' unique identifiers to them {\em before} users log in. (However, if the IdP and RPs collude {\em while} a user logs in, they can potentially identify users through timing side channels. Time-based attack is out of the scope of this paper and existing works~\cite{fett2015spresso, zhang2021passo}.)
\end{itemize}

\boldhead{Selective disclosure of user identity.} The user is given the ability to only disclose her pre-selected identity attributes that he or she wished to share with the RP. The user is supposed to hide her distinctive identity attributes such as email address and phone number to prevent account linkage across RPs in order to achieve full-scale privacy protection. With the assumption that there is no leakage outside the protocol, our protocol aims to allow the user to log in at the RP \emph{anonymously}, by revealing none of her identity attributes to the RP. 

\boldhead{Security requirements for the mixer.} Given an SSO instance, the mixer should not learn the identity of the involved user (note that the mixer can identify RPs and IdPs through their TLS public keys) nor can it impersonate the user on the RP.

\boldhead{Legacy-compatibility with OAuth 2.0.}
An IdP that supports OAuth 2.0 should be applied to our protocol \emph{without any modification}.
Also, RP only needs to present a button on its login page for initiating our protocol and perform the attestation check of the mixer upon the first use.
There are no special scripts or keys for the IdP and RPs.

This is the highlight feature of \name{} that distinguish is from works on cryptography-based privacy-preserving SSO solutions that either design new protocols or make changes to the OAuth 2.0 or OIDC protocol.
However, none of them achieve our privacy goals without modifying the OAuth 2.0 protocol of the IdP.
Since most commercial off-the-shelf IdPs are implemented over the OAuth 2.0 protocol, e.g., Google, Facebook, and Twitter, asking IdPs to adopt a modified OAuth 2.0 protocol or even an entirely different protocol is neither desirable nor practical.

\boldhead{Robust to single-point failures.} 
Our protocol aims to protect the user against single-point failures w.r.t. security and availability.
\emph{Security} is defined as: the attacker can not impersonate the user to log in to the RP by compromising some of her accounts on the IdPs.
\emph{Availability} is defined as: the user can still log in to her same account on the RP when some of the IdPs are unavailable, e.g., service downtime.

\section{\name{}}
\label{sec:design}

\begin{figure*}
  \centering
  \includegraphics[width=0.95\linewidth]{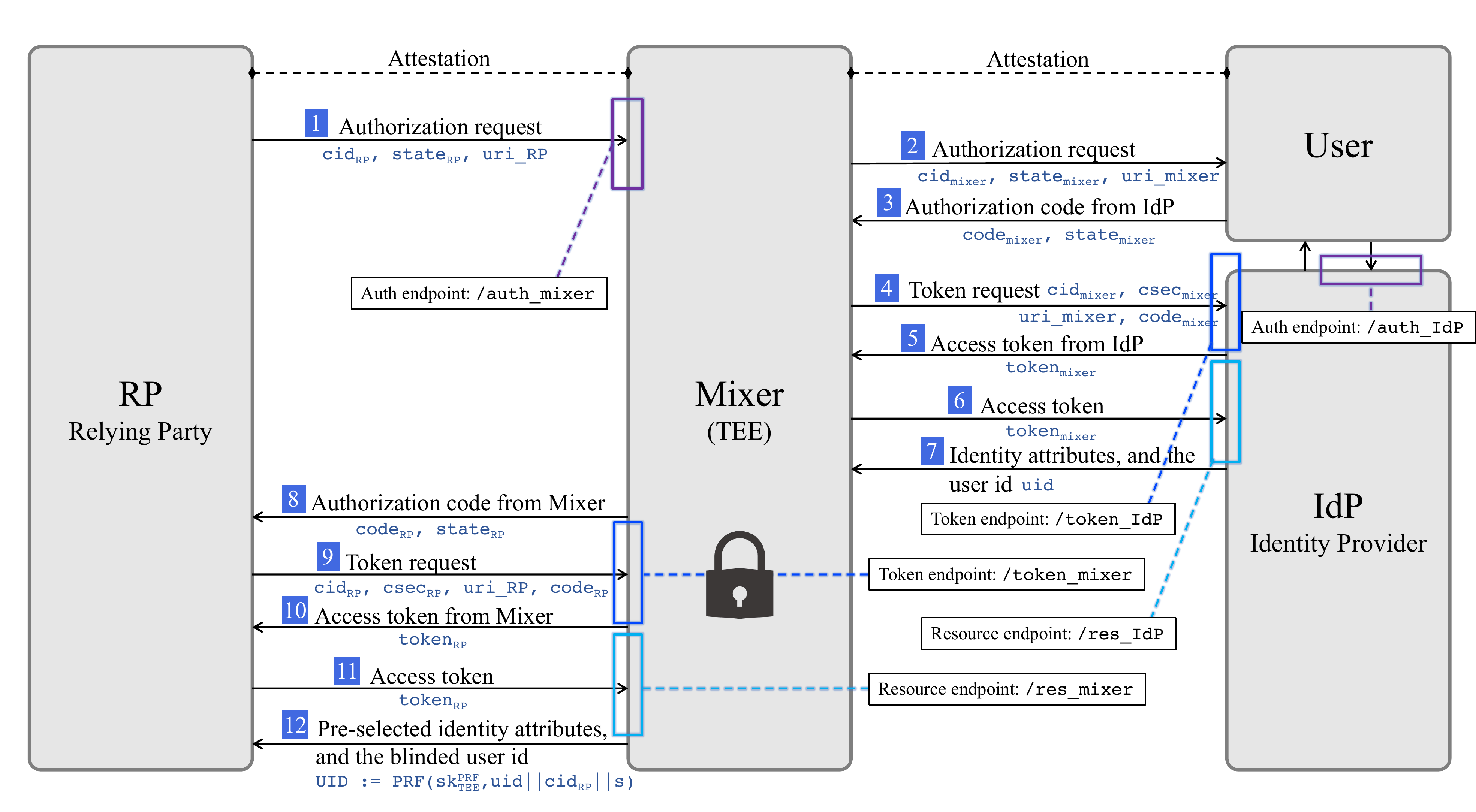}
  \caption{System architecture and the standard SSO workflow for \name{}.}
  \label{fig:design-abstract}
\end{figure*}

In this section, we present the system design of \name{} and specify the SSO workflow (\autoref{sec:basicprotocol}) and the formal \name{} protocol (\autoref{sec:formalprotocol} in message-driven fashion).
In \name{} architecture, there are four types of parties: \emph{user(s)} $\user$ (a user's \emph{user-agent} $\UA$ acts on behalf of her), the \emph{Relying Party(ies)} $\RP$, the \emph{Identity Provider(s)} $\IDP$, and the \emph{mixer} $\mixer$. 
Our system is designed to provide service for a large population of users and RPs $\RP_1,\cdots,\RP_m$ and IdPs $\IDP_1,\cdots,\IDP_k$. In general, a user may wish to use our system to sign-on $\RP_j$ with her privacy protected, and she authenticates with a certain $\IDP_i$, or a group of IdPs $\IDP_1, ..., \IDP_k$. In order to achieve privacy protection, $\RP_j$ communicates to the IdP(s) through $\mixer$ only.

\subsection{Overview of \name{}}
\label{sec:overview}

The main strategy of \name{} is that we manage all the interactions between $\RP$ and $\IDP$ through a central mixer $\mixer$, which is implemented with the Trusted Execution Environment (TEE).
\autoref{fig:design-abstract} illustrates the architecture of \name{} and the protocol workflow for SSO. 
At a high-level glance, the mixer $\mixer$ not only stands in the middle of the system but also breaks the system into two independent yet closely hinged parts.
From \autoref{fig:design-abstract}, if we leave $\RP$ aside and focus on the rest of the system, $\mixer$, $\user$, and $\IDP$ form a standard OAuth 2.0 trio, and the protocol flow is identical to that in \autoref{fig:oauth}. And if we leave $\user$ and $\IDP$ aside, $\RP$ and $\mixer$ make up an OAuth 2.0-like flow except that $\mixer$ makes the authorization grant on itself without the participation of the user $\user$. 

Therefore, the mixer $\mixer$ in \name{} \emph{mixes} the two nested OAuth 2.0 flows together by playing different roles towards different parties: $\mixer$ acts the role of a relying party when interacting with $\IDP$, and acts the role of an identity provider when interacting with $\RP$. 
Thus, from the perspective of $\IDP$, the $\mixer$ is no other than a normal relying party, and $\RP$ treats $\mixer$ as if it is a quasi-IdP with only minimal differences.
With this design, we are able to reuse the existing OAuth 2.0 protocol as much as possible to achieve the legacy-(OAuth 2.0)-compatibility.

The mixer $\mixer$ are required to be fully trusted, and keep the confidentiality and integrity of its code and data. 
\name{} leverages TEE to realize the functionality of $\mixer$.
The mixer $\mixer$ is a TLS server started in a TEE on a host. 
The user $\user$ and $\RP$ use remote attestation to make sure that it is indeed communicating with $\mixer$ secured by a TEE, initialized properly in the enclave.
Data inside $\mixer$ is inaccessible outside the enclave on the same host due to runtime isolation. Because of this, $\mixer$ can keep the secrets and keys privately. 
The sealing feature assures persistent storage for important credentials, so even if the mixer is down, long-term secrets sealed previously can be revived when the enclave restarts. 
In conclusion, we use TEE to ensure the trustworthiness of $\mixer$, which is the key to addressing our privacy threats stated.

\boldhead{Notation.}
In this and upcoming sections, we use letters in typewriter font, e.g., \texttt{$\mathtt{para}$}, to denote the parameters passed through the protocol. Note that unless we mention that a parameter is optional, it is treated as a MUST field. Other keywords ahead of a parameter are to be interpreted as described in RFC 2119 \cite{bradner1997key}. 
We use different subscripts ($\mathtt{A_{mixer}}$, $\mathtt{A_{RP}}$, and $\mathtt{A_{IdP}}$) to distinguish a similar type of parameters $\mathtt{A}$ that belong to different parties. 
E.g., \tokenB{} is the access token held by the mixer and \tokenRP{} is the access token held by the RP.
Similarly, we use the underscores (\texttt{B\_mixer} and \texttt{B\_IdP}) to distinguish a similar type of protocol endpoints or URIs \texttt{B}.

\subsection{\name{} Workflow}
\label{sec:basicprotocol}

Standard \name{} workflow consists of three parts, \emph{Mixer registration}, \emph{RP registration}, and \emph{login flow}. The two registration parts are the preparation processes before the login flow.
Before registration, $\mixer$ samples a $256$bit PRF key $\ktee{}$ and seals it to the disk.
$\user$ and $\RP$ need to verify the remote attestation of $\mixer$ if this is their \emph{first time} using \name{}. Attestation gives proof that the TEE is indeed executing the presumed code (the $\mixer$ code shall be open-sourced for public review). 
Also, prior to the login flow, the user $\user$ is supposed to set her preferences of selective-disclosure on the mixer $\mixer$ through a trusted channel such as submitting an HTTPS form, otherwise by default $\mixer$ will disclose none of her attributes to $\RP$, except a blinded user identifier.
\autoref{fig:design-abstract} illustrates the workflow (login flow).

\boldhead{Mixer registration.}
The mixer $\mixer$ registers with $\IDP$ as an OAuth 2.0 relying party, i.e., acts as a relying party from $\IDP$'s perspective. The parameters below are the same as that in the OAuth 2.0 standard. $\mixer$ provides its redirect URI \redirecturiB{} and other required information depending on the certain specification of $\IDP$, and gets its credentials, the client identifier \clientidB{} and the client secret \clientsecB{} from $\IDP$.
In this workflow, $\mixer$ needs to register at all the identity providers which it wishes to support, and saves its credentials in persistent storage for future use ($\mixer$ actually uses private keys generated in TEE to \emph{seal} data inside the TEE to the encrypted disk).
    
\boldhead{RP registration.} 
$\RP$ registers with $\mixer$. This is almost identical to the RP registration in OAuth 2.0 except $\RP$ registers with $\mixer$ instead of $\IDP$. This time, $\mixer$ acts as an identity provider from the $\RP$'s perspective. $\RP$ must submit its redirect URI \redirecturiRP{} to get its credentials, the client identifier \clientidRP{}, and the client secret \clientsecRP{} from $\mixer$. 
For each RP that participates in this workflow, it only needs to register with $\mixer$ once. 
$\mixer$ saves the client credentials of $\RP$ securely for future authentication, and the redirect URI \redirecturiRP{} with the corresponding $\RP$'s client credentials.

\boldhead{Login flow.}
This workflow is constructed from two OAuth 2.0 authorization code grant flows. Hence, we denote the OAuth 2.0 flow which consists of $\RP$, $\mixer$ as the \emph{outer OAuth 2.0}\footnote{This is not a standard OAuth 2.0 workflow since $\user$ is absent, but except for this, the flow is identical to an OAuth 2.0 flow. Thus it does not break the legacy-compatibility.} since both the beginning and the finishing steps of the login flow are contained in this OAuth 2.0 flow. Similarly, we denote the workflow which consists of $\mixer$, $\IDP$, and $\user$ as the \emph{inner OAuth 2.0}, as this flow is executed in the middle of the login flow. $\mixer$ has three protocol endpoints to serve resources to $\RP$:

\begin{itemize}
    \item \emph{Authorization endpoint (\emph{\authendB{}}).} This is used by $\RP$ to obtain the authorization code \codeRP{} from $\mixer$, via user-agent $\UA$'s redirection.
    \item \emph{Token endpoint (\emph{\tokenendB{}}).} This is used by $\RP$ to exchange the authorization code \codeRP{} for the access token \tokenRP{}, after $\RP$ authenticates with $\mixer$ successfully with its credentials \clientidRP{}, \clientsecRP{}.
    \item \emph{Resource endpoint\footnote{This endpoint is not specified in the OAuth 2.0 framework.} (\emph{\resourceendB{}}).} This is used by $\RP$ to retrieve user $\user$'s identity attributes from $\mixer$ with its access token. 
\end{itemize}
$\IDP$ also owns three protocol endpoints to offer similar services but to $\mixer$, we denote them with \authendIdP{}, \tokenendIdP{}, and \resourceendIdP{}.
The usages, forms, and specifications of the authorization codes \codeRP{}, \codeB{}, and the access tokens \tokenRP{}, \tokenB{} follow their descriptions in the OAuth 2.0 framework, readers may refer to \autoref{sec:codentoken} for a brief overview, but beware that the \codeRP{} is issued from $\mixer$ to $\RP$ and the \codeB{} is issued from $\IDP$ to $\mixer$, same interpretations for the access tokens.

The login flow has a \emph{sandwich}-like structure, both logically and chronologically.
It begins with the outer OAuth 2.0 workflow, then steps into the inner OAuth 2.0, and returns to the outer OAuth 2.0 again when the inner workflow finishes. Before $\RP$ initiates the login flow, the user $\user$'s user-agent $\UA$ loads the login page of $\RP$. 

We specify the login flow as follows, the indexing of the steps is consistent with that in \autoref{fig:design-abstract}:

\begin{enumerate}

\item[\mysquare{1}] ($\RP \stackrel{\rm \UA}{\longrightarrow} \mixer$) This is $\RP$'s authorization request: $\RP$ initiates the flow by directing $\user$'s user-agent $\UA$ to $\mixer$'s authorization endpoint \authendB{} with its client identifier \clientidRP{}, a local state \localstateRP{} (a random generated nonce to prevent CSRF attacks) and the redirect URI \redirecturiRP{} to which $\mixer$ will redirect $\UA$ back later after the inner OAuth 2.0 workflow ends.

\item[\mysquare{2}] ($\mixer \stackrel{\rm \UA}{\longrightarrow} \IDP$) This is $\mixer$'s authorization request. After $\mixer$ checks that the client identifier \clientidRP{} and redirect URI \redirecturiRP{} are consistent with those $\RP$ got issued in the \emph{RP registration} step, $\mixer$ then directs $\UA$ to the $\IDP$'s authorization endpoint \authendIdP{} with its client identifier \clientidB{}, a local state \localstateB{} and the redirect URI \redirecturiB{} for the future redirection to $\mixer$. $\IDP$ authenticates $\user$ via $\UA$, e.g., by checking $\user$'s account and password on $\IDP$, and then establishes whether $\user$ grants or denies the authorization request (made by $\RP$ and transmitted by $\mixer$) via a consent window. If $\user$ denies, the flow terminates. If $\user$ grants, the flow goes on.

\item[\mysquare{3}] ($\IDP \stackrel{\rm \UA}{\longrightarrow} \mixer$) Assuming $\user$ grants the request, $\IDP$ redirects $\UA$ back to $\mixer$ with its redirect URI \redirecturiB{} provided earlier. The redirection includes the authorization code \codeB{} and any local state \localstateB{} passed by $\mixer$ earlier.
    
\item[\mysquare{4}] ($\mixer \to \IDP$) $\mixer$ requests the access token \tokenB{} from $\IDP$'s token endpoint \tokenendIdP{} by including the authorization code \codeB{} received in the previous step. When making the request, $\mixer$ authenticates with $\IDP$ using the \clientidB{} and \clientsecB{}. The redirect URI \redirecturiB{} used to obtain the authorization code \codeB{} should also be included for verification.

\item[\mysquare{5}] ($\IDP \to \mixer$) $\IDP$ authenticates $\mixer$ and validates its authorization code \codeB{}, and ensures that the redirect URI \redirecturiB{} received matches the URI used for redirection. If valid, $\IDP$ responds with the access token \tokenB{}.

\item[\mysquare{6}] ($\mixer \to \IDP$) $\mixer$ uses its access token \tokenB{} to retrieve $\user$'s unique user identifier \userid{} and other user attributes from $\IDP$'s resource endpoint \resourceendIdP{}. The user identifier \userid{} is unique among all users on the certain IdP and shall never be reused. 

\item[\mysquare{7}] ($\IDP \to \mixer$) $\IDP$ returns the \userid{} along with other attributes.
If $\user$ has selected to disclose any attributes previously, $\mixer$ keeps these attributes. 
$\mixer$ calculates a \emph{pre}-blinded user identifier, dubbed \preuid{}:
\[
\preuid{} \coloneqq \prf{}(\ktee{}, \userid{} \| \clientidRP{})
\]
$\mixer$ generates a per-user salt $\mathsf{s}$ and saved to the encrypted disk in a table $(\preuid,\mathsf{s})$.
\preuid{} is used as the query of the per-user salt.
Finally, $\mixer$ calculates the blinded user identifier \uid{}:
\[
\uid{} \coloneqq \prf{}(\ktee{}, \userid{} \| \clientidRP{} \| \mathsf{s})
\]
The \uid{} is the \emph{only} user identifier that is passed to $\RP$ for recognizing $\user$.
    
\item[\mysquare{8}] ($\mixer \stackrel{\rm \UA}{\longrightarrow} \RP$) If the inner OAuth 2.0 flow succeeds, $\mixer$ redirects $\UA$ back to $\RP$ with its redirect URI \redirecturiRP{} provided earlier. The redirection includes the authorization code \codeRP{}{} and any local state \localstateRP{} passed by $\RP$ earlier.

\item[\mysquare{9}] ($\RP \to \mixer$) $\RP$ requests the access token \tokenRP{} from $\mixer$'s token endpoint \tokenendB{} by including the authorization code \codeRP{} received in the previous step. When making the request, $\RP$ authenticates with $\mixer$ using the \clientidRP{} and \clientsecRP{}. The redirect URI \redirecturiRP{} used to obtain the authorization code \codeRP{} should also be included for verification.

\item[\mysquare{10}] ($\mixer \to \RP$) $\mixer$ authenticates $\RP$ and validates its authorization code \codeRP{}, and ensures that the redirect URI \redirecturiRP{} received matches the URI used for redirection. If valid, $\mixer$ responds with the access token \tokenRP{}.

\item[\mysquare{11}] ($\RP \to \mixer$) $\RP$ uses its access token \tokenRP{} to retrieve $\user$'s \uid{} from $\mixer$'s resource endpoint \resourceendB{}. $\RP$ can also use the \tokenRP{} to retrieve the attributes $\user$ has already selected to disclose.

\item[\mysquare{12}] ($\mixer \to \RP$) $\mixer$ returns \uid{} along with other attributes (if any).
        
\end{enumerate}

The result of the successful execution of the login flow is that $\RP$ obtains the user $\user$'s unique user identifier \uid{} from $\mixer$. $\RP$ is supposed to keep a list of unique user identifiers \uid{}s for a certain $\IDP$. Once the \uid{} it received is off the list, it should create a new profile for $\user$ and add it to the list. Next time, $\user$ will be authenticated successfully if $\RP$ gets the same \uid{} again.

\begin{figure*}
    \centering

\begin{boxedminipage}[t]{\textwidth}

\begin{center}
    {$\PROT_{\name} \text{ between } \IDP, \RP \text{ and }\mixer$}
\end{center}
\vspace{1pt}
 
 \begin{pchstack}[center]
\procedure[mode=text, linenumbering, codesize=\footnotesize]{}{
\underline{$\PROT_{\IDP}$}:\\
follow the standard OAuth 2.0 protocol (see \autoref{fig:protocolIdP}). \\[1mm]
\underline{$\PROT_{\RP}$}: \\
(\textbf{RP registration})\\
\onrecv{} ($\msg{register}$) from environment:\\
\quad \cmt{verify attestation first}\\
\quad send ($\msg{get-att}$) to $\mixer$ \\
\quad wait to receive ($\pktls, \sigtee$)\\
\quad $\pktee = \gatt\mathsf{.getpk}()$\\
\quad assert $\sigschemetee \mathsf{.Vf}(\pktee, \sigtee, \pktls)$ \\
\quad send ($\msg{register}, \redirecturiRP{}$) to $\mixer$ \\
\quad wait to receive ($\clientidRP, \clientsecRP$) \\
\quad save ($\clientidRP, \clientsecRP$) to disk\\
(\textbf{Login flow})\\
follow the standard OAuth 2.0 protocol (see \autoref{fig:protocolRP}). \\[1mm]
\underline{$\PROT_{\mixer}$}: \\
(\textbf{Init})\\ 
\oninit{}: \\
\quad send ($\msg{install}, \program $) to $\gatt$ \cmt{$\program$ is $\mixer$'s code}\\
\quad send ($\eid, \msg{resume}$) to $\gatt$\\
\normalsize{\underline{$\program $}}: \scriptsize\cmt{$\mixer$'s code is now running in $\gatt$}\\
\oninit{}: \\
\t $\ktee  \sample \bin^{256}$ \cmt{the PRF key}\\
\t seal $\ktee$ to disk\\
\onrecv{} ($\msg{get-att}$): \\
\t load $\pktls$ and compute $\pktls$ \\
\t $\sktee = \gatt\mathsf{.getsk}()$ \\
\t $\sigtee =\sigschemetee \mathsf{.Sig}(\sktee, \pktls)$ \\
\t \pcreturn $(\pktls, \sigtee)$ \\
(All messages below are received through a TLS channel \\
established using $\pktls$.) \\[2mm]
(\textbf{RP registration})\\
\onrecv{} ($\msg{register}, \redirecturiRP{}$) from $\RP$: \\
\quad check if $\redirecturiRP{}$ is a valid address\\
\quad let $\randseed \sample \bin^{256}$\\
\quad let $\clientidRP \coloneqq \hash (\randseed \| \redirecturiRP)$\\
\quad sample $\clientsecRP \sample \bin^{256}$ \\
\quad seal $(\clientidRP, (\redirecturiRP, \clientsecRP))$ to disk\\
\quad \pcreturn ($\clientidRP, \clientsecRP$)
}
\procedure[mode=text, codesize=\footnotesize, lnstart=39, linenumbering]{}{
\normalsize{\underline{$\program $}} \scriptsize{(Cont'd)}: \\
(\textbf{Mixer registration})\\ 
\onrecv{} ($\msg{register}$) from environment:\\
\quad register with $\IDP$ using $\redirecturiB$\\
\quad receive and seal ($\clientidB{}, \clientsecB{}$)\\
(\textbf{Login flow})\\
\onrecv{} ($\msg{auth},\clientidRP, \localstateRP, \redirecturiRP$) from $\RP$:\\
\quad query $\redirecturiRP'$ using $\clientidRP$  \\
\quad assert $\redirecturiRP' = \redirecturiRP$\\
\quad select $\codeB$ (by OAuth 2.0 standard) \cmt{authorization code}\\
\quad sample $\localstateB{} \sample \ZZ$ \cmt{random local state}\\
\quad send ($\msg{auth}, \clientidB, \localstateB, \redirecturiB$) to $\IDP$\\
\onrecv{} ($\msg{auth}, \codeB, \localstateB'$) from $\IDP$:\\
\quad assert $\localstateB' = \localstateB$\\
\quad send ($\msg{token}, \clientidB, \clientsecB, \redirecturiB, \codeB$) to $\IDP$\\
\onrecv{} ($\msg{token}, \tokenB$) from $\IDP$:\\
\quad send ($\msg{identity}, \tokenB$) to $\IDP$\\
\quad wait to receive $\userid$ \\ 
\quad sample $\mathsf{s} \sample \bin^{256}$ \\
\quad \cmt{assume the output length of the $\prf{}$ is 256 bits}\\
\quad let $\preuid{} \coloneqq \prf{}(\ktee{}, \userid{} \| \clientidRP{})$ \\
\quad seal $(\preuid,\mathsf{s})$ to disk \\
\label{line:uidcalc} \quad let $\uid{} \coloneqq \prf{}(\ktee{}, \userid{} \| \clientidRP{} \| \mathsf{s})$ \\
\quad \cmt{keep the $\uid$ in the memory is enough, do NOT seal it to disk}\\
\quad send ($\msg{auth}, \codeRP, \localstateRP$) to $\RP$\\
\onrecv{} ($\msg{token}, \clientidRP, \clientsecRP, \redirecturiRP, \codeRP$) from $\RP$:\\
\quad \textbf{If} $\clientsecRP, \redirecturiRP, \codeRP$ match with $\clientidRP$ \textbf{then}\\
\quad \quad select $\tokenRP$ (by OAuth 2.0 standard) \cmt{access token}\\
\quad \quad send ($\msg{token}, \tokenRP$) to $\RP$\\
\quad \textbf{else} output failure\\
\onrecv{} ($\msg{identity}, \tokenRP'$) from $\RP$:\\
\quad assert $\tokenRP' = \tokenRP$\\
\quad \pcreturn $\uid$ \\
\quad \cmt{by default $\mixer$ discloses no user attributes other than }\\
\quad \cmt{a blinded \uid{} to $\RP$}
}
\end{pchstack}
\end{boxedminipage}
    \caption{The \name{} protocol}
    \label{fig:protocol}
\end{figure*}

\subsection{\name{} Protocol}
\label{sec:formalprotocol}

We formally specify the \name{} protocol in \autoref{fig:protocol}.
To formalize the attested execution model of TEE, we adopt the ideal functionality $\gatt$ proposed by Pass \etal \cite{cryptoeprint:2016/1027}. We can model TEE's group signature as a regular signature scheme $\sigschemetee$, which contains a pair of public and secret keys, and denoted $\pktee$ and $\sktee$.
In an attested execution on the TEE, a party first send a user-defined program $\program$ inside a new enclave, dubbed the \msg{install} call. Upon installation, $\gatt$ generates an enclave and returns $\eid$, the enclave identifier that can be used to identify the enclave instance. Upon a \msg{resume} call with the $\eid$, $\gatt$ executes the program $\program$ over the inputs $\inputtee$, obtaining an output $\outputtee$ and a signature $\sigtee = \sigschemetee \mathsf{.Sig}(\sktee,(\program,\outputtee))$ ($\sktee = \gatt\mathsf{.getsk}()$). The signature $\sigtee$ is the resulting attestation of the enclave. 
To verify an attestation, the verifier party need to use $\pktee$, which can be obtained with $\gatt\mathsf{.getpk}()$.

The \name{} protocol specification is presented in $\PROT_{\name}$ (\autoref{fig:protocol}) (supplementary protocols are given in \autoref{sec:pseudo}). The protocol relies on TEE's ideal functionality $\gatt$. Readers can combine $\PROT_{\name}$ with \autoref{fig:design-abstract} and \autoref{sec:basicprotocol} for a clearer comprehension.

\subsection{Transitioning to Multi-IdP SSO}

The \name{} architecture shows its strong capabilities in building complex authentication schemes. We demonstrate its potential with the \extendedprotocol{} workflow, under the extended setting of multiple IdPs.
Similar to the concept of multi-signature, which allows several private keys to jointly generate a single signature, the \extendedprotocol{} workflow allows users to use identities from multiple IdPs to log in to a single RP. Also, our workflow supports \mofn{2}{3} schemes. 
Details of the extended version of \name{} is deferred to \autoref{sec:extended}.

\section{Security Analysis}
\label{sec:analysis}

\subsection{Security and Privacy Properties}
\label{subsec:analysis}

Under the assumptions outlined in~\autoref{subsec:adversarial-model}, we will show that \name{} satisfies the security and privacy goals outlined in \autoref{subsec:goals}.
At a high level, the security and privacy of \name{} boils down to that of TLS and TEE. 
The mixer starts a TLS server in a TEE. Upon initialization, it will generate a pair of TLS keys, publish the public key with an attestation proving its correctness (e.g., as done in~\cite{zhang2016town,cheng2019ekiden}) while keeping the secret key in the enclave. The user (via the user-agent) verifies the remote attestation in a trust-upon-first-use manner to verify that the given TLS public key is securely generated by a TEE so that the corresponding secret key is kept secret. 
Then, the user can establish a secure channel with the mixer TEE.
Using TLS ensures that the mixer host cannot learn user identities other than the IP addresses. 
In \autoref{tab:leakage}, we list the information learned by each party in MISO and the standard SSO protocol.

\begin{table}
\centering
\caption{Information leakage to each party.}
\begin{threeparttable}
\begin{tabular}{cp{2.3cm}p{2.5cm}}
\toprule
\diagbox{Party}{Protocol} &MISO &SSO (OAuth 2.0) \\
\midrule
IdP &mixer's client id \clientidB  & RP's client id at the IdP, RP's IP address \\
RP &blinded user id \uid  &user id \userid, user's attribute at the IdP, e.g., email, user's IP address, IdP's IP address \\
mixer &user's IP address, IdP's IP address, RP's IP address &--- \\
\bottomrule
\end{tabular}
\end{threeparttable}
\footnotesize
\label{tab:leakage}
\end{table}

\boldhead{Unlinkability.} We will prove that \name{} not only achieves IdP unlinkability and RP unlinkability, but also maintains unlinkability even in the case of collusion between IdP and RPs.

\begin{figure}
    \centering
\begin{pcimage}
    \createprocedureblock{procb}{center,boxed}{}{}{linenumbering, width=\linewidth}
    \procb{$\mathsf{G}^{unlink}_{\name{}, \IDP{}}(\secparam, \adv)$}{
    \text{initialize $\gatt{}[\program]$} \\
    IdP \gets \adv^{\gatt{}[\program]}\\
    \text{simulate an user $U$; register $U$ at $IdP$ \cmt{\adv{} gets $\userid{}$}} \\
    \text{simulate two RPs ($RP_{0}$, $RP_{1}$)}\\
    \text{register $RP_{0}$, $RP_{1}$ at $\gatt{}[\program]$} \\
    b \sample \bin \\
    \text{$U$ login to $RP_{b}$ following $\PROT_{\name}$} \\
    b' \sample \adv{}() \\
    \pcreturn b=b'
    }
    \end{pcimage}
    \caption{IdP unlinkability game.}
    \label{fig:idpunlinkgame}
\end{figure}

\begin{figure}
    \centering
    \begin{pcimage}
    \createprocedureblock{procb}{center,boxed}{}{}{linenumbering, width=\linewidth}
    \procb{$\mathsf{G}^{unlink}_{\name{}, \RP{}}(\secparam, \adv)$}{
    \text{initialize $\gatt{}[\program]$} \\
    RP_0, RP_1 \gets \adv^{\gatt{}[\program]} \\
    \text{simulate the $IdP$ and two user ($U_0$, $U_1$)} \\
    \text{$U_0$, $U_1$ login to $RP_{0}$ following $\PROT_{\name}$}\\
    \cmt{\adv{} gets $\uid_{0, 0}$, $\uid_{0, 1}$, $\uid{}_{i,j}$ is the $\uid{}$ of $U_j$ on $RP_i$} \\
    b \sample \bin \\
    \text{$U_b$ login to $RP_1$ following $\PROT_{\name}$} \\
    b' \sample \adv(\uid_{1, b}) \\
    \pcreturn b=b'
    }
    \end{pcimage}
    \caption{RP unlinkability game.}
    \label{fig:rpunlinkgame}
\end{figure}

\begin{figure}
    \centering
    \begin{pcimage}
    \createprocedureblock{procb}{center,boxed}{}{}{linenumbering, width=\linewidth}
    \procb{$\mathsf{G}^{unlink}_{\name{}}(\secparam, \adv)$}{
    \text{initialize $\gatt{}[\program]$} \\
    IdP, RP_{0}, RP_{1} \gets \adv^{\gatt{}[\program]} \\
    \text{simulate two users ($U_0$, $U_1$); register $U_0$, $U_1$ at $IdP$} \\
    \cmt{\adv{} gets $\userid{}_0$, $\userid{}_1$ of $U_0$, $U_1$} \\
    \text{$U_0$, $U_1$ login to $RP_{0}$ following $\PROT_{\name}$}\\
    \cmt{\adv{} gets $\uid_{0, 0}$, $\uid_{0, 1}$, $\uid{}_{i,j}$ is the $\uid{}$ of $U_j$ on $RP_i$} \\
    b \sample \bin \\
    \text{$U_b$ login to $RP_1$ following $\PROT_{\name}$} \\
    b' \sample \adv(\uid_{1, b}) \\
    \pcreturn b=b'
    }
    \end{pcimage}
    \caption{Collusive-IdP-RP unlinkability game.}
    \label{fig:unlinkgame}
\end{figure}

\begin{definition}[IdP Unlinkability]
\name{} offers IdP unlinkability if, for any stateful $\ppt$ adversary $\adv$, $\prob{\mathsf{G}^{unlink}_{\name{}, \IDP{}}(\lambda, \adv) \Rightarrow 1} \leq \frac{1}{2} + \negl[\lambda]$.
~\autoref{fig:idpunlinkgame} specifies the game $\mathsf{G}^{unlink}_{\name{}, \IDP{}}$, in which the adversary \adv{} is the honest-but-curious IdP. The challenger simulates an user $U$ and registers $U$ with the IdP, allowing \adv{} to learn $U$'s \userid{}. Next, the challenger simulates two RPs $RP_0$ and $RP_1$, and registers them at the mixer (specifically the enclave $\gatt{}[\program]$). The challenger picks a random bit $b\in\bin$ and simulate the login flow with $\IDP{}$ and $RP_b$. The adversary must guess which RP is picked.
\label{def:idpunlink}
\end{definition}

\begin{definition}[RP Unlinkability]
\name{} offers RP unlinkability if, for any stateful $\ppt$ adversary $\adv$, $\prob{\mathsf{G}^{unlink}_{\name{}, \RP{}}(\lambda, \adv) \Rightarrow 1} \leq \frac{1}{2} + \negl[\lambda]$.
~\autoref{fig:rpunlinkgame} specifies the game $\mathsf{G}^{unlink}_{\name{}, \RP{}}$, in which the adversary controls two RPs, $RP_0$ and $RP_1$. The challenger simulates two users $U_0$ and $U_1$ and logs $U_0$ and $U_1$ in to $RP_0$, allowing \adv{} to learn $U_0$ and $U_1$'s $\uid{}_{0,0}$ and $\uid{}_{0,1}$ on $RP_0$.
Then, the challenger randomly picks a user $U_b$ to login $RP_1$, and the adversary must guess which user is picked.
\label{def:rpunlink}
\end{definition}

\begin{definition}[Collusive-IdP-RP Unlinkability]

\name{} offers collusive-IdP-RP unlinkability if, for any stateful $\ppt$ adversary $\adv$, $\prob{\mathsf{G}^{unlink}_{\name{}}(\lambda, \adv) \Rightarrow 1} \leq \frac{1}{2} + \negl[\lambda]$.
~\autoref{fig:unlinkgame} specifies the game $\mathsf{G}^{unlink}_{\name{}}$, in which the adversary controls two RPs, $RP_0$ and $RP_1$, and can get $\userid{}$s of the users on the IdP. The challenger simulates two users $U_0$ and $U_1$, and registers them with the IdP, allowing \adv{} to learn the $\userid{}_0$ and $\userid{}_1$ of $U_0$ and $U_1$. Subsequently, the challenger enables both  $U_0$ and $U_1$ to log in to $RP_0$ using the prescribed $\PROT_{\name}$.
Then, the challenger randomly picks a user $U_b$ to login $RP_1$, and the adversary must guess which user is picked. 
Note that compared to~\autoref{def:rpunlink}, $\adv$ learns not only $\uid{}_{i,j}$ but also $\userid{}_j$ for each  $RP_i$ and user $U_j$.

\label{def:idprpunlink}
\end{definition}

We note that IdP unlinkability and RP unlinkability are weaker guarantees than collusive-IdP-RP unlinkability. IdP unlinkability specifies that the IdP cannot determine if a user has logged into an RP, while RP unlinkability specifies that given a \uid{} on the RP, the RP cannot determine if a user has logged in or not. The definition of collusive-IdP-RP  expresses the infeasibility of collusive IdP and RPs to determine whether a user logs in an RP or not. We prove that \name{} guarantees collusive-IdP-RP unlinkability as follows.
\begin{proof}

The confidentiality provided by TEE guarantees that the adversary cannot get the PRF key ($\ktee{}$).

In the setup phase, according to the power of adversary specified in the $\mathsf{G}^{unlink}_{\name{}}$, the adversary \adv{} can obtain $\userid{}_0$ and $\userid{}_1$ of $U_0$ and $U_1$, $\clientidRP{}_0$ and $\clientidRP{}_1$ of $RP_0$ and $RP_1$, as well as the $\uid{}_{0, 0}$ and $\uid{}_{0, 1}$ subsequent to the users' login to $RP_0$.
Then, upon $U_b$ logging into $RP_1$, the adversary \adv{} can acquire $\uid{}_{1, b}$.

According to the $\PROT_{\name}$ (line~\#\autoref{line:uidcalc}), $\uid{}_{1, b}$ is calculated from $ \mathsf{PRF}(\ktee{}, \userid{}_{b} \| \clientidRP{}_{1} \| \mathsf{s})$. To prove the unlinkability in \name{}, it is equivalent to prove that, given $\userid{}_i$ and $\uid{}_{0, i}$ for $i \in \bin$ and $\uid{}_{1,b}$, the adversary \adv{} cannot determine $b$.

We prove this using a hybrid argument. Initially, suppose there is a $\mixer{}'$ that substitutes the \prf{} with a random function (the resulting system is denoted as $\name{}'$). If $\uid{}_{1, b}$ is computed from $\userid{}_{b}$ by a random function, then the probability of \adv{} winning the game is equivalent to bit guessing, i.e., $\prob{\mathsf{G}^{unlink}_{\name{}'}(\lambda, \adv) \Rightarrow 1} = \frac{1}{2}$.

Next, we will show that \adv{} cannot distinguish $\mathsf{G}^{unlink}_{\name{}}$ and $\mathsf{G}^{unlink}_{\name{}'}$.
If \adv{} can distinguish between $\mathsf{G}^{unlink}_{\name{}}$ and $\mathsf{G}^{unlink}_{\name{}'}$ with an advantage $\advantage{}{} = \left| \prob{\mathsf{G}^{unlink}_{\name{}} \Rightarrow 1} - \prob{\mathsf{G}^{unlink}_{\name{}'} \Rightarrow 1} \right|$, we can construct an adversary \bdv{} with the same advantage to win the PRF game\cite{boneh2023graduate} (see ~\autoref{fig:prfattackgame}) by following:
\bdv{} submits a sequence of queries in the form of $x_i \coloneqq \userid{}_i\|\clientidRP{}\|s$ ($i = 1, 2, \cdots$) to the challenger. The challenger randomly selects a function $f$ to computes $y_i \gets f(x_i)$, which will then be provided to \bdv{}. Subsequently, \bdv{} will forward the sequence of results $y_i$ to \adv{} and pass each $y_i$ off as $\uid{}_{i}$. Finally, \bdv{} will provide the output of \adv{} to the challenger as its own output.

Let $W_b$ be the event that \bdv{} outputs 1 in the PRF game for $b \in \bin$, the advantage of \bdv{} is $\advantage{\prf{}}{\bdv{}} = \left| \prob{W_0} - \prob{W_1} \right|$. The reduction (see  ~\autoref{fig:misoreductiongame}) shows $\prob{W_0} = \prob{\mathsf{G}^{unlink}_{\name{}} \Rightarrow 1}$ and $\prob{W_1} = \prob{\mathsf{G}^{unlink}_{\name{}'} \Rightarrow 1}$. $\advantage{\prf{}}{\bdv{}} = \left| \prob{W_0} - \prob{W_1} \right| = \left| \prob{\mathsf{G}^{unlink}_{\name{}} \Rightarrow 1} - \prob{\mathsf{G}^{unlink}_{\name{}'} \Rightarrow 1} \right| = \advantage{}{}$.
Thus, we have successfully created an adversary \bdv{} who can win the PRF game with $\advantage{}{}$. This implies that \adv{} cannot distinguish between $\mathsf{G}^{unlink}_{\name{}}$ and $\mathsf{G}^{unlink}_{\name{}'}$, except for a  negligible probability. Given that $\prob{\mathsf{G}^{unlink}_{\name{}'}(\lambda, \adv) \Rightarrow 1} = \frac{1}{2}$, we have $\prob{\mathsf{G}^{unlink}_{\name{}}(\lambda, \adv) \Rightarrow 1} \leq \frac{1}{2} + \negl[\lambda]$.
\qedhere
\end{proof}

As per our proof, \name{} not only guarantees collusive-IdP-RP unlinkability but also IdP unlinkability and RP unlinkability.

\boldhead{Selective disclosure.} 
This property follows immediately from the integrity and confidentiality of TEEs.
In \name{}, users specify the scope of information to be given to an RP. The mixer will enforce the specification (integrity) and intercept and drop data that users wish to keep secret (confidentiality).

\boldhead{Security of the mixer.} 
Due to the confidentiality guarantee of TEEs and the security of TLS, the mixer (the malicious host) cannot infer the identity of the user from encrypted traffic. (We assume users connect to the mixer via anonymity network to hide network-level identities.) Due to the integrity of TEEs, the mixer will only emit a token to RP after successfully authenticating the user with the IdP, thus the mixer cannot impersonate users.

\subsection{Tolerating Side-channel Leakage}

TEEs are well-known to be subject to side-channel leakage~\cite{van2018foreshadow, lee2017hacking, murdock2020plundervolt, chen2019sgxpectre} that can break confidentiality. 
We take three measures to mitigate the impact of potential side channels, in addition to standard measures such as using constant-time cryptography implementation. %

\boldhead{Using salt to slow down rainbow-table attacks.}
In the case of TEE compromise, the PRF key $\ktee{}$ will be leaked. Then, if a strong adversary choose to collude with the malicious RP, he could launch rainbow-table attacks~\cite{hellman1980cryptanalytic,oechslin2003making} to recover \userid{} from the blinded \uid{}. 
We slow down such attacks by constructing \uid{} with a \emph{per-user} salt (both in basic protocol and \extendedprotocol{}).
The salt table is also kept in the same disk with $\ktee{}$.
Even if the salts are leaked, the adversary has to mount attacks by constructing \emph{per-salt} rainbow-tables, which significantly slows down the attack.

\boldhead{Keeping secrets short-lived.} Secret values such as the access tokens obtained from IdPs (\tokenB{}) only reside in the memory for a short period of time, and we remove them from the memory as soon as the SSO protocol concludes.
With this token, the attacker can query the IdP for private user information (up to the scope set by the user when authenticating to the IdP).
However, by keeping secrets short-lived, they are less likely to be leaked through side-channel attacks (though that is not guaranteed). 

\boldhead{Rotating long-term secrets.} Some secrets live in the memory for a relatively long period of time, with the most crucial one being the TLS key. 
If the attacker can obtain the TLS key, then she can decrypt the traffic and forge messages, and all bets are off. To combat, one can frequently rotate the TLS key, which however requires users and RPs to verify attestations each time. A better approach is to leverage hierarchical attestation~\cite{zhang2017rem} by running the attestation verification logic in another TEE $\text{TEE}_\text{verify}$. This way, users and RPs only need to verify one attestation produced by $\text{TEE}_\text{verify}$ (which establishes the public key of $\text{TEE}_\text{verify}$), after which they can accept any public key signed by $\text{TEE}_\text{verify}$ with regular certificate chain verification, providing a better user experience.

\subsection{Limitations of \name{}}

\boldhead{IdP-mixer collusion.} If the IdP colludes with the malicious host of the mixer, they can break the unlinkability by IdP since the host of the mixer can identify the RP and IdP connected to the mixer through their TLS public keys. This does not require breaking the TEEs. In practice, however, forming such collusion require a degree of trust between an IdP and the mixer, which can be hard to establish (in particular, most IdPs are big tech firms who will incur a reputation damage if such collusion is exposed). %

\boldhead{Account linkage from network-level identities.}
We note that malicious RPs can achieve linkability across RPs through users' IP addresses. To preserve unlinkability across RPs, users need to use an anonymity network such as mix networks \cite{chaum1981untraceable} or Tor \cite{dingledine2004tor}. 
If the user fails to use such networks, a curious mixer host can also link certain IP addresses with users. 
Even in this case, compared with the basic SSO (e.g., Google), a curious mixer in \name{} still tracks \emph{much less} user information, e.g., the mixer can only link network-level identities (not other sensitive user attributes like emails).

\section{Implementation and Evaluation}
\label{sec:implement}

In this section, we present implementation and performance evaluation.

\pgfplotsset{
every tick label/.append style={scale=0.75},
every axis/.append style={
  }
}

\begin{figure}
\centering
    \begin{tikzpicture}
    
    \begin{axis}[
        width=1\linewidth,
        height=6.5cm,
        title={},
        xlabel={Concurrent users$/$sec},
        x label style={scale=0.9},
        ylabel={Avg. latency (ms)},
        y label style={at={(1.3,0.5)},anchor=north,scale=0.9},
        xmin=-50, xmax=1050,
        ymin=0, ymax=3700,
        xtick={10,200,400,600,800,1000},
        ytick={0,500,1000,1500,2000,2500,3000,3500,4000},
        yticklabel pos=right,
        legend cell align=left,
        legend pos=north west,
        legend style={nodes={scale=0.6, transform shape}}, 
        ymajorgrids=true,
    ]
        
    \addplot[
        color=myblue!50,
        mark=o,
        mark options={solid,mark size=1.5pt},
        line width=1.1pt,
        error bars/.cd, y dir=both,y explicit
        ]
        coordinates {
        (10,34) +- (2.8,2.8)
        (200,406) +- (38,38)
        (400,743) +- (67,67)
        (600,1092) +- (71,71)
        (800,1457) +- (128,128)
        (1000,1932) +- (149,149)
        };
        \addlegendentry{standard SSO w.o. SGX}
        
    \addplot[
        color=myblue!90,
        mark=*,
        mark options={solid,mark size=1.5pt},
        line width=1.1pt,
        error bars/.cd, y dir=both,y explicit
        ]
        coordinates {
        (10,37) +- (2.6,2.6)
        (200,454) +- (43,43)
        (400,889) +- (104,104)
        (600,1309) +- (129,129)
        (800,1632) +- (171,171)
        (1000,2201) +- (125,125)
        };
        \addlegendentry{standard SSO}
        
     \addplot[
        color=mydeepblue!50,
        mark=triangle,
        mark options={solid},
        line width=1.1pt,
        error bars/.cd, y dir=both,y explicit
        ]
        coordinates {
        (10,43) +- (3.3,3.3)
        (200,634) +- (67,67)
        (400,1110) +- (109,109)
        (600,1651) +- (152,152)
        (800,2117) +- (160,160)
        (1000,2753) +- (236,236)
        };
        \addlegendentry{\mofn{2}{3} \extendedprotocol{} w.o. SGX}
        
    \addplot[
        color=mydeepblue!90,
        mark=triangle*,
        line width=1.1pt,
        error bars/.cd, y dir=both,y explicit
        ]
        coordinates {
        (10,56) +- (4.3,4.3)
        (200,773) +- (79,79)
        (400,1365) +- (101,101)
        (600,1823) +- (124,124)
        (800,2382) +- (198,198)
        (1000,3108) +- (273,273)
        };
        \addlegendentry{\mofn{2}{3} \extendedprotocol{}}
        
    \end{axis}
    \end{tikzpicture}
\caption{Concurrency is shown in average login time. Each user sends one login request.}
\label{fig:performance}
\end{figure}

\subsection{Prototype Implementation}

To demonstrate the capability of \name{} under different use cases, we implement a prototype that works under standard SSO authentication and \extendedprotocol{} authentication. 
The core task of prototype implementation is the mixer, which is implemented in an Intel SGX enclave.

We choose the open-sourced EGo SDK \cite{edgeless2022effortlessly} to implement the mixer. EGo is a LibOS (Library Operating System)-based SGX SDK that supports the execution of Golang applications in SGX. A LibOS SGX SDK \emph{packs} the important functions in the OS kernel into libraries and packs into an enclave directly, so as to avoid handling the interface between the OS and enclave manually.
A great feature of EGo is that it provides easy-to-use APIs for remote attestation (to host a TLS server inside the enclave, a common approach is to let the enclave generate the TLS keys itself and bind it to a remote attestation statement).
We use Golang's native TLS package \texttt{crypto/tls}\footnote{\url{https://pkg.go.dev/crypto/tls}} to provide services at the two protocol endpoints.
As a TLS server, the mixer needs a set of certificates for client verification. Since the enclave should not trust the host's root certificate, we obtain X.509 certificates that issued from \texttt{Certbot}\footnote{\url{https://certbot.eff.org/pages/about}}. We provide these certificates to the mixer by embedding (also a feature supported by EGo) them in the enclave binary.
We use the \texttt{prf12}\footnote{\url{https://go.dev/src/crypto/tls/prf.go\#L66}} from package \texttt{crypto/tls} for our $\mathsf{PRF}$ function.
Since the mixer plays different roles for the two types of OAuth 2.0 flows, and we use the OAuth 2.0 client-side package \texttt{golang.org/x/oauth2}\footnote{\url{https://pkg.go.dev/golang.org/x/oauth2}} for the mixer implementation as an OAuth 2.0 relying party, we use the OAuth 2.0 server-side package \texttt{go-oauth2/oauth2/v4}\footnote{\url{https://pkg.go.dev/github.com/go-oauth2/oauth2/v4}} for the mixer implementation as an OAuth 2.0 identity provider. For persistent storage of long-term secrets of the mixer, we use the package \texttt{ego/ecrypto}\footnote{\url{https://pkg.go.dev/github.com/edgelesssys/ego/ecrypto}} for sealing keys and credentials inside the enclave. 

We implement two mixer-side enclaves to support the standard SSO workflow and a \mofn{2}{3} \extendedprotocol{} workflow in about 1.7k lines of Golang code in total.
To demonstrate the full usage scenarios, we also implement an RP\footnote{Technically, the only modification required by the RP is to display a button on its login page to allow the user to trigger SSO with \name{}.} that supports OAuth 2.0 standard in Golang.
For both the mixer and the RP, we build web pages for the demo presentation.
We use commercial off-the-shelf IdPs, in particular, Google, Facebook, and Github.
There is no specific requirement for user-agents, and we have tested the prototype with Google Chrome and Apple Safari.
Demo of our prototype is presented in \autoref{a:demo}.

\begin{figure}
    \centering
    \begin{tikzpicture}
    \begin{axis}[
    	width=1.06\linewidth,
        height=6.5cm,
        ybar,
        xlabel={Concurrent users$/$sec},
        x label style={scale=0.9},
        ylabel={Avg. latency (ms)},
        y label style={at={(1.25,0.5)},anchor=north,scale=0.8},
        ymin=0, ymax=2000,
        xtick=data,
        ytick={0,500,1000,1500,2000},
        yticklabel pos=right,
        yticklabel style={scale=0.8},
    	enlarge x limits=0.2,
        enlarge y limits={value=.06,upper},
        legend cell align=left,
    	legend style={at={(0.26,0.96)},nodes= {scale = 0.6, transform shape},anchor=north,legend columns=1},     
        nodes near coords,
        nodes near coords style={scale=0.51,color=black!60},
        nodes near coords align={vertical},
    ]
    \addplot [draw=plum,fill=plum!30] 
    	coordinates {(200,207) (400,388) (600,570) (800,853) };
    \addplot [draw=myburgundy!60,fill=myburgundy!30]
    	coordinates {(200,454) (400,889) (600,1309) (800,1783) };
    \addplot [draw=myburgundy,fill=myburgundy!60]
    	coordinates {(200,443) (400,867) (600,1274) (800,1770) };
    \legend{SSO (OAuth 2.0), \name{} SSO (MRENCLAVE), \name{} SSO (MRSIGNER)}
    \end{axis}
    \end{tikzpicture}
    \caption{Comparing \name{} with basic SSO}
    \label{fig:performance-added}
\end{figure}

\subsection{Performance Evaluation}

To evaluate the performance of \name{} prototype, we pose the following research questions, two general ones and two SGX related ones:
\begin{enumerate}
    \item[\textbf{\req{1}.}] What is the scalability of \name{} (the standard and the Multi-IdP one) under increasing user concurrency?
    \item[\textbf{\req{2}.}] How long is the latency introduced in standard \name{} compared with a basic (without the mixer) SSO?
    \item[\textbf{\req{3}.}] How big is the overhead caused by the use of SGX?
    \item[\textbf{\req{4}.}] What are the performance differences between different SGX sealing modes, i.e. MRENCLAVE and MRSIGNER?
\end{enumerate}

\boldhead{Experiment setup.} We deploy the mixer enclaves on a server running Ubuntu 22.04.2 LTS on an Intel $^\circledR$ Xeon $^\circledR$ E-2274G CPU with 30GB RAM. 
In order to avoid the variance of network latency between the IdP and the mixer, we set up all the servers including the mixer, the IdP, and the RP on the same machine.
For simplicity, we assume a selective disclosure policy where the user shares none of her identity attributes with the RP.

The latency we tested is the end-to-end time from the beginning to the end of the user login, i.e., the user login time.
We use the Apache Jmeter\texttrademark{} \cite{apachejmeter} to create concurrent users' log in requests. The level of \emph{concurrency} is the number of different users trying to perform SSO login per second. 
We simulate different levels of concurrency with the duration of each test for $1$ minute.

\boldhead{Answer to \req{1}.} To answer \req{1}, we deploy \emph{two} enclaves of the mixer, one for the standard SSO workflow and one for the \mofn{2}{3} \extendedprotocol{} workflow. The heap size for both enclaves is adjusted to 4096MB. We measure the latency and the standard deviation, the results are shown in \autoref{fig:performance}. 
Results suggest our prototype can handle about $1000$ concurrent users. At $1000$ concurrent users$/$sec, the prototype provides a latency of $2.20$ seconds for the standard SSO and $3.11$ seconds for phase II of the \mofn{2}{3} \extendedprotocol{}.

\boldhead{Answer to \req{2}.} To answer \req{2}, we compare the standard \name{} SSO with the basic SSO (OAuth 2.0 protocol) without the mixer under the same codebase. The performance comparison is presented in \autoref{fig:performance-added}. Results suggest \name{} roughly incurs $2 \times$ more latency when comparing with the basic one --- This is in line with our intuition because \name{} nests $2$ (basic) SSO workflows at a higher level. 

\boldhead{Answer to \req{3}.} To answer \req{3}, we run tests in which the mixer is not implemented with SGX as (insecure) baselines to demonstrate the overhead introduced by SGX.
The results are shown in \autoref{fig:performance}. 
We use $\overline{T}$ to denote the average latency \emph{under all concurrency} with a specific SSO prototype, i.e., the standard one or the \mofn{2}{3} \extendedprotocol{}. We use $\overline{T}_{baseline}$ to denote that without SGX (insecure baselines). The overhead caused by SGX $O_{\text{SGX}}$ is calculated as:
\[
    O_{\text{SGX}} = \frac{(\overline{T} - \overline{T}_{baseline})}{\overline{T}_{baseline}} \times 100\%
\]
For \req{3}, the overhead for the standard \name{} SSO is $14.67\%$, and that for the phase II of the \mofn{2}{3} \name{} SSO is $13.91\%$.

\boldhead{Answer to \req{4}.} SGX supports two sealing modes: MRENCLAVE mode that allows only the enclave binary itself to decrypt the encrypted data, and MRSIGNER mode that allows the binary signer to decrypt. 
The performance of these two sealing modes is also presented in \autoref{fig:performance-added}. 
The performance difference is \emph{minimal}. This is mainly because there are not many operations involving sealing in the login workflow, so its overhead is insignificant relative to network latency and CPU fluctuations.
We use the MRENCLAVE mode by default\footnote{MRSIGNER mode enables the creator of the binary to decrypt the encrypted disk, which can cause problems if the creator is also compromised. Readers can consult~\cite[Page 21]{intel2018guide} for more details on how to choose between the two sealing modes.}.

\boldhead{Comparison with previous work.} 
It is difficult to give a rigorous performance comparison with most related works, since the experiment setups are different (e.g., EL PASSO \cite{zhang2021passo} deploy their IdP and RP on cloud, but the user-agent on a local device, we deploy the IdP, RP, and user-agent on a same server), thus the roundtrip latency involved in the protocols are also hard to estimate \cite{fett2015spresso, zhang2021passo, guo2021uppresso}.
We compare two aspects of \name{} with other related works: concurrency and latency.
Firstly, \name{} is more scalable under high user concurrency (below $800$). We notice that in EL PASSO, the latency deteriorates seriously as the concurrency exceeds $200$ (see ~\cite[Figure 7]{zhang2021passo}).
We only find UPPRESSO \cite{guo2021uppresso} enjoys the similar experiment setup with \name{}.
The latency of UPPRESSO is $2.84 \times$ over basic SSO without mentioning of concurrency. Their experiment also suggests that the latency of SPRESSO \cite{fett2015spresso} is $3.02 \times$ over basic SSO.
In comparison with these two works, \name{} has a shorter latency of $2.19 \times$ over basic SSO under $200$ concurrency.
In conclusion, \name{} achieves \emph{shorter latency and higher scalability} compared with previous works.

In a real-world deployment, the overhead of network communication between users, IdPs, RPs, and the mixer could grow larger due to their distances. However, this cost is incurred by all deployed SSO systems.
In addition, the mixer can be distributed across multiple SGX instances on more servers if a higher concurrency is desired.

\section{Discussion}
\label{sec:discussion}

In this section we discuss the compatibility of \name{} with OAuth 2.0 implicit grant flow, the OIDC protocol, and authorization. Afterwards we point out future extensions for our system.

\boldhead{Compatibility with the implicit grant flow.}
Although our protocol specified~\autoref{fig:protocol} is built over OAuth 2.0 authorization grant flow, \name{} is also compatible with the \emph{implicit grant} flow \cite[\S 4.2]{hardt2012rfc}.
When working with the implicit grant flow in the inner OAuth 2.0, the mixer can directly obtain \tokenB{} from the response of the authorization request instead of \codeB{}. The steps during which the mixer exchanges \codeB{} for the \tokenB{} are omitted (steps \mysquare{4} \mysquare{5} in \autoref{fig:design-abstract}).
Similarly, when working with an RP in the outer OAuth 2.0, the mixer issues \tokenRP{} as a hash fragment to the user-agent, then let the user-agent pass it directly to the web-based RP, and steps \mysquare{9} \mysquare{10} in \autoref{fig:design-abstract} are omitted.

However, we discriminate the implicit grant flow from the authorization grant flow because it offers \emph{less security} guarantee, and it is vulnerable to user impersonation attacks \cite[\S 10.16]{hardt2012rfc}.
Thus, the usage of this flow should be limited to web-based, public RPs (see~\autoref{subsec:oauthprotocol}) that are unable to authenticate themselves with the IdP, e.g., native applications installed on the user's device \cite[\S 9]{hardt2012rfc}.

\boldhead{Compatibility with OIDC.}
\name{} is compatible with OIDC.
In OIDC protocol, an IdP issues an \emph{ID token} along with the access token in response to the token request, see \autoref{fig:oauth} step \mysquare{4}.
An ID token is a JSON Web Token (JWT)~\cite{rfc7519} signed by the issuer IdP, in which concludes claims about the user's identity information (e.g., user identifier and email). 
Therefore, the RP can obtain the user identifier directly from the ID token in OIDC, without retrieving it from the IdP using the access token as in OAuth 2.0.
The \userid{} in our protocol is identical to the \texttt{sub} field of an ID token \cite[\S 2]{oidcfinal}.
When working with OIDC, the protocol steps in which the mixer retrieves user identifier and other attributes are omitted (steps \mysquare{6} \mysquare{7} \mysquare{10} \mysquare{11} in \autoref{fig:design-abstract}).
Our prototype supports Google (OIDC-based) as the IdP, demonstrating this compatibility.

\boldhead{Privacy-preserving authorization.}
\emph{Authorization}, the ability to empower a third party (RP) to act on behalf of the user to require resources from the IdP (a.k.a. resource server), is supported by OAuth 2.0 from the beginning. 
Hence, \name{} adapts to authorization without modifying the protocol workflow.
However, the mixer may need to manage more fine-grained privacy policies for the user when the resources required by the RP conflict with privacy-preserving goals.

\boldhead{The mixer's involvement in \name{}.}
An malicious RP may wish to redirect the user-agent to the IdP directly instead of the mixer to go through a normal SSO. However, if the RP does this, the authentication step (step \mysquare{2} in \autoref{fig:design-abstract}) in later interactions would be \emph{distinguishable} from the user.
The user can clearly notice that the malicious RP wants to obtain the user's authorization instead of the mixer in a consent window like \autoref{fig:demo-google}, i.e., it would appears to be ``If you continue, ... with \emph{`MyApp'}'' instead. Then, the user can simply deny the consent to terminate this SSO instance.

\boldhead{Future works.} 
One possible extension is to make use of multiple TEEs (potentially heterogeneous ones from different vendors, such as ARM TrustZone \cite{alves2004trustzone, winter2008trusted}, AMD Secure Encrypted Virtualization (SEV) \cite{amdsev}, GPU TEEs \cite{volos2018graviton, nvidiatee}, and Keystone (RISC-V) \cite{keystone}) to further decentralized the trust the mixer. MPC and hybrid approaches (e.g., combining TEE with MPC) are also interesting future directions to explore.
\section{Related Work}
\label{sec:related}

\begin{table*}
\caption{Privacy-preserving SSO and anonymous credentials solutions.}

\centering\fontsize{8}{10}\selectfont

\newcommand*\emptycirc[1][0.7ex]{\tikz\draw (0,0) circle (#1);} 
\newcommand*\halfcirc[1][0.7ex]{%
  \begin{tikzpicture}
  \draw[fill] (0,0)-- (90:#1) arc (90:270:#1) -- cycle ;
  \draw (0,0) circle (#1);
  \end{tikzpicture}}
\newcommand*\fullcirc[1][0.7ex]{\tikz\fill (0,0) circle (#1);} 

\begin{threeparttable}
\begin{tabularx}{\textwidth}{c*9{>{\centering\arraybackslash}X}}
\toprule
 &\multicolumn{3}{c}{\textbf{SSO Properties}}
 &\multicolumn{4}{c}{\textbf{Privacy-preserving Features}}
 &\multicolumn{2}{c}{\textbf{Others}}\\
\diagbox{\textbf{Solution}}{\textbf{Feature}}
&User Identification at the RP
&User Authentication only at the IdP 
&IdP-confirmed Identity Retrieval 
&IdP Unlinkability 
&RP Unlinkability
&Collusive-IdP-RP Unlinkability
&Selective Disclosure 
&Browser-only \tnote{1}
&Legacy Compatibility\\
\midrule
\textbf{Privacy-preserving SSO} \\
OIDC with PPID\cite{sakimura2014openid} &\cmark &\cmark &\cmark &\xmark &\cmark &\xmark &\halfcirc \tnote{5} &\cmark &\cmark\\
Mozilla BrowserID \cite{fett2015analyzing} &\cmark &\halfcirc \tnote{2} &\xmark &\cmark &\xmark &\xmark &\xmark &\cmark &\xmark\\
PseudoID \cite{dey2010pseudoid} &\cmark &\xmark &\halfcirc \tnote{3} &\cmark &\cmark &\cmark &\cmark &\cmark &\xmark\\
SPRESSO \cite{fett2015spresso} &\cmark &\cmark &\halfcirc \tnote{3} &\cmark &\xmark &\xmark &\halfcirc \tnote{5} &\cmark &\xmark\\
PRIMA \cite{asghar2018prima} &\cmark &\xmark &\cmark &\cmark &\xmark &\xmark &\cmark &\xmark &\xmark\\
Sign in with Apple \cite{signinwithapple} &\cmark &\cmark &\cmark &\xmark &\cmark &\xmark &\cmark &\cmark &\halfcirc \tnote{6}\\
POIDC\tnote{7} \cite{hammann2020privacy} &\cmark &\cmark &\cmark &\cmark &\cmark &\xmark &\halfcirc \tnote{5} &\cmark &\xmark \tnote{6}\\
EL PASSO \cite{zhang2021passo} &\cmark &\xmark &\cmark &\cmark &\cmark &\cmark &\cmark &\cmark &\xmark\\ 
UPPRESSO \cite{guo2021uppresso} &\cmark &\cmark &\cmark &\cmark &\cmark &\xmark &\cmark &\cmark &\xmark \tnote{6}\\
\hline
\rowcolor{grayhighlight} \name{} &\cmark &\cmark &\cmark &\cmark &\cmark &\cmark &\cmark &\cmark &\cmark \\
\hline
\textbf{Anonymous Credentials} \\
U-prove \cite{paquin2011uprovetech,paquin2011uprovecrypto} &\xmark &\xmark &\xmark &\cmark &\cmark &\cmark &\cmark &\xmark &\xmark\\
ABC4Trust \cite{baigneres2014d4, rannenberg2015attribute} &\cmark &\xmark &\cmark &\cmark &\cmark &\cmark &\cmark &\xmark &\xmark\\
UnlimitID \cite{isaakidis2016unlimitid} &\cmark &\xmark &\xmark &\halfcirc \tnote{4} &\cmark &\xmark &\cmark &\xmark &\xmark\\
IRMA \cite{alpar2017irma} &\xmark &\xmark &\xmark &\cmark &\cmark &\cmark &\cmark &\xmark &\xmark\\
Fabric Idemix \cite{hyperledgerfabric} &\xmark &\xmark &\xmark &\cmark &\cmark &\cmark &\cmark &\xmark &\xmark\\
\bottomrule
\end{tabularx}
\end{threeparttable}
\begin{tablenotes}
\footnotesize
    \item \halfcirc{} denotes a feature that is partially supported by a solution.
    \item{1} The user only needs a browser to finish the whole SSO process. She does not need to provide certain hardware or install specific software.
    \item{2} The user of BrowserID keeps a private key to sign a token namely the identity assertion, and the signature needs to be verified by the RP.
    \item{3} PseudoID uses zero-knowledge proofs to provide blindly-signed user attributes, but not implemented. SPRESSO can be extended to provide user attributes in tokens, but has not been implemented.
    \item{4} The user blinds its credential attributes and deposits it to the IdP’s authorization endpoint, which gives the possibility of account linkage with respect to the IdP.
    \item{5} OIDC and OIDC-based POIDC only allow the user to select disclose some of her attributes (the user cannot disclose none of her attributes). SPRESSO can be extended to selective disclosure of user attributes in tokens, but has not been implemented.
    \item{6} OIDC-based POIDC and UPPRESSO can only compatible to the implicit grant flow with modifications on the IdP. OAuth 2.0-based Sign-in with Apple only supports Apple as the IdP.
    \item{7} POIDC is not implemented.
\end{tablenotes}
\label{tab:sso_solutions}
\end{table*}

Our paper is generally related to four lines of works: 1) privacy-preserving SSO schemes that augment single sign-on workflow with privacy guarantees; 2) anonymous credentials that achieve similar privacy goals but in a different model (i.e., they are not single sign-on); 3) decentralized identity schemes which bear similar ideas to our \extendedprotocol{} protocol design; 4) works that also leverage TEE to address authentication and authorization-related problems.

\subsection{Privacy-preserving SSO and Anonymous Credentials}

We classify prior and the most related works on privacy-preserving SSO and anonymous credential schemes in \autoref{tab:sso_solutions}. The major difference is that the latter does not satisfy SSO properties as defined in~\autoref{sec:SSO properties}.

\boldhead{Privacy-preserving SSO.} OIDC core 1.0 \cite{sakimura2014openid} specifies a privacy protection mechanism, namely the Pairwise Pseudonymous Identifier (PPID) to provide RP unlinkability. When using the PPID feature, the IdP issues an identifier (PPID) that identifies the user to the RP, which cannot be correlated with the user's PPID at another RP. However, the IdP records a user's PPIDs for a certain RP, thus violates IdP unlinkability. 
The Mozilla BrowserID (a.k.a. Persona) \cite{fett2015analyzing} is the very first commercial SSO solution that aims for privacy protection. However, severe security attacks like malicious RP impersonation attacks are discovered by Fett \etal in \cite{fett2014expressive} and it is no longer supported in the market.
PseudoID \cite{dey2010pseudoid}, a privacy-preserving SSO solution based on the OpenID \cite{recordon2006openid}, utilizes blind signatures. The user is required to blind a token and then present it to the IdP. Later the IdP asserts the token and lets the user log in to the RP. The blind signed token contains no RP identity, and such tokens are unlinkable by the RPs. Thus unlinkability towards IdP and RP are both achieved. However, in this system, the user uses her secrets to log in to the RP, which adds extra steps to the login process.
Fett \etal proposed SPRESSO \cite{fett2015spresso}, in which they achieved IdP unlinkability. Their solution also breaks the direct communications between the IdP and the RP with a client-side agent called the forwarder. The RP uses a pseudo-identity for each login to hide its identity. However, the IdP returns the same user identifier to the RP each time, which results in RP unlinkability. 
In PRIMA \cite{asghar2018prima}, a solution built over Oblivion \cite{simeonovski2015oblivion}, decouples the communications between the IdP and the RP. The IdP signs the credentials with a key and user's attributes. The user uses the key to convince the RP about her identity. This solution gives much burden to its user in comparison with a purely SSO scheme. It does not provide RP unlinkability.
Apple's Sign in with Apple service \cite{signinwithapple} can cope with private email relay \cite{signinwithappleemailrelay} to allow the user to use an anonymous email address to log in to RPs. Thus, the user's email address is protected, and she can still receive emails from the RP with the relay service. Apple's solution can not provide IdP unlinkability unless the user fully trust Apple. Furthermore, it limits users to use Apple as the only IdP choice and relies on their own infrastructure and software.
Hammann \etal proposed the Privacy-Preserving OpenID Connect (POIDC) with the pairwise POIDC \cite{hammann2020privacy}, two protocol extensions for OIDC. The pairwise POIDC offers both IdP unlinkability and RP unlinkability. However, it involves a zero-knowledge proof (ZKP) between the parties that brings more cost. They have not implemented their protocols so the performance is not given. Li and Mitchell point out POIDC is unpractical in \cite{li2020user}.
Grassi \etal proposed the Privacy-Enhanced Identity Brokers \cite{grassi2015privacy}, which introduce a similar broker design to achieve the privacy goals. However, they don not specify the protocol and are unclear about how to achieve the trustworthiness of the broker.
El PASSO \cite{zhang2021passo} is a privacy-preserving \emph{asynchronous} SSO solution, in which the user obtains her anonymous credentials from the IdP and keeps them. When logs in to the RP, she proves the ownership of her credentials and selectively discloses her attributes. However, the user is required to be authenticated by the RP during each login, which is kind of against the original wishes of \emph{single} sign-on. 
UPPRESSO \cite{guo2021uppresso} provides both IdP and RP unlinkability. During each login, the user and the RP negotiate a pseudo-identity and send it to the IdP, later the IdP returns the user's pseudo-identity to the RP. However, it requires modification for the IdP and only supports the OIDC implicit grant flow.

\boldhead{Anonymous credentials.}
Anonymous credentials systems allow users to use a secret key to log in to the RP. However, most of such solutions do not support SSO features defined in \autoref{sec:SSO properties}, not to mention the OAuth 2.0 legacy compatibility.
We list popular solutions in \autoref{tab:sso_solutions} because they are still sound solutions for user privacy protection.
Systems like U-prove \cite{paquin2011uprovecrypto,paquin2011uprovetech}, ABC4Trust \cite{baigneres2014d4, rannenberg2015attribute}, UnlimitID \cite{isaakidis2016unlimitid}, IRMA \cite{alpar2017irma}, and Hyperledger Fabric Idemix \cite{hyperledgerfabric} all provides IdP and RP unlinkability (IdP unlinkability is partially supported in UnlimitID).
The biggest drawback of these systems is that the user needs to take care of the secret key in such systems.
Most anonymous credentials solutions require specific user-side installations for credentials management and also suffer from noticeable user-perceived overheads for computing the anonymous credentials. 
Thus, anonymous credentials systems are seen as less adopted in the market in comparison with prominent SSO protocols such as OAuth 2.0 and OIDC.
In conclusion of \autoref{tab:sso_solutions}, \name{} is the very first OAuth 2.0-compatible privacy-preserving SSO solution.

\subsection{Decentralized Identity}
Decentralized identity, a.k.a. self-sovereign identity, is an identity management scheme that allows the user to self-control her identity without relying on a centralized IdP.
Unlike traditional identity management systems, in which exists a specific, centralized, and trusted entity (IdP) takes charge of the issuance, control, and maintenance and control of a user's identity. 
Initiatives such as the W3C Decentralized Identifiers (DIDs) work group \cite{w3cdecentralized} and Decentralized Identity Foundation \cite{decentralizedfoundation} have been developing standards to support decentralized identity.
However, an obstacle to the large adoption of such DID systems \cite{hyperledgerindy, microsoftdid, veramo} is that they fail to leverage users' identities available through existing web services, such as conventional centralized and federated identity systems. 
In \name{}, our \extendedprotocol{} workflow is legacy-{IdP}-compatible, which supports users to port identities from existing SSO-based identity services, e.g., Google, Facebook, and privacy-preserving as well.
Maram \etal proposed CanDID \cite{maram2021candid}, a platform for the practical, user-friendly, and privacy-preserving realization of DID. Their approach is also legacy-compatible but different from ours: they leverage an oracle, e.g., DECO \cite{zhang2020deco} or Town Crier \cite{zhang2016town} to port identities from existing web services.
Torus \cite{torus} is a secure passwordless authentication and private key management platform which integrates OpenLogin~\cite{openlogin} (now web3auth~\cite{web3auth}). The OpenLogin is an authentication scheme also facilitates OAuth 2.0 standards. However, their approach does not offer as many privacy-preserving features as \name{} does, e.g., the initial login requests a user’s minimal public information, users can not selectively disclose the information at the IdP side. Also, like Apple, Torus only claims but not guarantees that they log the user out of her 3rd party account immediately after her identity is verified.

\subsection{TEE-based Authentication and Authorization Systems}

TEEs are widely adopted for many scenarios. For example, Intel SGX, in particular, has been utilized in various applications, including high performance state-machine replication protocol \cite{behl2017hybrids}, secure Linux containers \cite{arnautov2016scone}, secure and privacy-enhancing Tor \cite{kim2017enhancing}, and decentralized private web search extensions \cite{pires2018cyclosa}.
To the best of our knowledge, there is no available study to address privacy-preserving SSO or federated identity solutions using TEE. Few works show TEE's capabilities in authentication and authorization. 
Zhang \etal proposed Town Crier \cite{zhang2016town}, which provides authenticated data feed for smart contracts. Town Crier serves trusted off-chain data from existing web services to smart contracts. Despite Town Crier runs a full web session inside a TEE, using Town Crier to implement similar SSO tasks completed by MISO requires no less effort than implementing MISO itself.
Matetic \etal proposed DelegaTEE \cite{matetic2018delegatee}, which leverages TEE to achieve delegation. DelegaTEE works mainly for sharing credentials between the user and third parties. Although it can be modified to perform the authorisation task similar to SSO. A key technical difference is that MISO does not require the user to deposit her credentials on the mixer/TEE, which makes MISO more robust to side-channel leakages.
\section{Conclusion}
\label{sec:conclusion}

In this paper, we present the first legacy-(OAuth 2.0)-compatible privacy-preserving SSO authentication architecture, namely \name{}. 
We enforce TEEs for realizing the system.
\name{} achieves four security and privacy goals: user account unlinkability by the IdP, unlinkability across RPs, selective disclosure of user identity, and robust to single-point failure. Our solution requires no modifications to commercial off-the-shelf IdPs. 
\name{} works with two authentication workflows, SSO workflow under the setting of a single IdP and the \extendedprotocol{} workflow under a setting of multiple IdPs, which demonstrates the potential and flexibility of our system.
We provide security analysis to our system, which shows that \name{} preserves security and privacy against powerful attackers. 
Our prototype implementation and evaluation suggest that \name{} enjoys high usability in real-world applications.

\boldhead{Positioning.} We believe that \name{} can significantly enhance security and privacy for a wide range of Internet users due to the large deployment of the SSO authentication infrastructure.
Most importantly, \name{} is highly practical since it is the first OAuth 2.0-compatible privacy-preserving SSO solution to date.

\section*{Acknowledgements}
We would like to thank the anonymous reviewers for their valuable comments that improved this paper, and Thomas Tendyck from EGo for answering questions about the EGo environment.
Zhixuan Fang and Fan Zhang are corresponding authors of this paper.

\bibliographystyle{IEEEtran}
\bibliography{IEEEabrv,ref}

\begin{appendices}
\section{OAuth 2.0 Backgrounds}
\label{sec:oauthbackground}

\subsection{Parties}
In the scenario of SSO, we reform OAuth 2.0 to a three-party protocol, and under this setting: 
\begin{itemize}
    \item The \emph{User} holds her credentials on the IdP. In most circumstances, the \emph{user-agent} acts on behalf of the user.
    \item The \emph{Relying Party (RP)} is a server to which the user logs in. Third-party applications are the RPs in SSO. 
    \item The \emph{Identity Provider (IdP)}, which is the joint entity of the authorization server and the resource server in OAuth 2.0, keeps the identity attributes of the user and issues the \emph{access token} to the RP to enable it to acquire those attributes. 
\end{itemize}

\subsection{Protocol}
\label{subsec:oauthprotocol}

The abstract protocol flow of OAuth 2.0 is illustrated in \autoref{fig:oauth}. In short, protocol starts by the \emph{authorization request} sent from the RP in step (1). If the user grants the request and returns the \emph{authorization grant} as response in step (2), the RP makes the \emph{token request} by sending the authorization grant the IdP in step (3). Then the IdP checks and issues the \emph{access token} in step (4). Thus, the RP is enabled to retrieve the user's identity attributes from the IdP using the access token in step (5) and (6).

Before initiating the protocol flow, the RP is required to register at the IdP to obtain its credentials on the IdP, we refer this step to \emph{RP Registration}. OAuth 2.0 framework classifies the RPs into two types, \emph{confidential} RPs which are capable for maintaining the confidentiality its credentials, e.g., client implemented on a secure server and has the ability to be authenticated, \emph{public} RPs which behave on the contrary. 
In this paper, we only focus on the confidential RPs since most web application servers have the ability to keep credentials securely. 
During the registration process, the RP hands in its redirect URI \emph{$redirect\_uri$} for user-agent redirection when the IdP returns the authorization grant in \autoref{fig:oauth} step (2). In return, a client identifier \emph{$client\_id$} and a client secret \emph{$client\_secret$} used for RP authentication between \autoref{fig:oauth} step (3) and (4) are issued by the IdP.

The significance of OAuth 2.0 framework is for the RP to gain the access token from the IdP so as to retrieve the attributes of the user. Based on that, the framework provides four main distinct set of detailed protocol flows.
In this paper, we focus on the specific protocol namely the \emph{authorization code grant} flow, since the implicit grant flow are specified for public RPs and the other two protocols are incapable under SSO scenario. The authorization code grant protocol got it name for the authorization grant in \autoref{fig:oauth} step (2) and (3) takes the form of an \emph{authorization code}. 
In the rest of this paper, we shorten the OAuth 2.0 authorization code grant protocol to OAuth 2.0 protocol.
More details on this protocol, we recommend readers to refer to \cite[\S 4.1]{hardt2012rfc}.

\subsection{Code and Token}
\label{sec:codentoken}
Below we introduce the two type of tokens (the authorization code and the access token) in OAuth 2.0.

\begin{itemize}
    \item \emph{Authorization code}: This is an opaque string, which is bounded to the identifier \emph{$client\_id$} and the redirect URL \emph{$redirect\_uri$} of an RP. The usage of an authorization code in OAuth 2.0 protocol is for the RP to exchange the access token from the IdP.
    \item \emph{Access token}: This is an opaque string, which is used to authorize RP's access to the user's identity attributes stored at the IdP. This token is encoded by the IdP and then issued to the RP, which contains the scope and duration. It takes the form of bearer token \cite{jones2012oauth}, which suggests any party can make use of it once obtained.
\end{itemize}

\subsection{Privacy Issues}

As far as we know, OAuth 2.0 protocol poses user's privacy concerns from various aspects. One is the \emph{$client\_id$} generated at the RP Registration stage and used frequently through the whole protocol flow, such as the authorization request and the token request, which result in the account linkage with respect to the IdP.
The other is the access token used for user's attributes retrieving. The attributes contains user's identity which the RP can access, leads to the user identity exposure. 
Furthermore, since the IdP keep the same user identifier to different RPs for the same user, which causes the account linkage across RPs. 
The OIDC protocol, which has largely adopted OAuth 2.0 protocol for its basis, making those privacy vulnerabilities hold true.

\section{\name{} Supplementary Protocols}
\label{sec:pseudo}

Protocol supplements for $\PROT_{\name}$ are specified in \autoref{fig:protocolIdP} and \autoref{fig:protocolRP}.
They are both \emph{standard} OAuth 2.0 protocol without any modification.

\begin{figure}
    \centering

\begin{boxedminipage}{\linewidth}

\begin{center}
    \scriptsize{$\PROT_{\IDP}$ (Login flow)}
\end{center}

\procedure[mode=text, codesize=\footnotesize,linenumbering]{}{
\onrecv{} ($\msg{auth}, \clientidB, \localstateB, \redirecturiB$) \\from $\mixer$:\\
\quad query $\redirecturiB'$ using $\clientidB$  \\
\quad assert $\redirecturiB' = \redirecturiB$\\
\quad send an authentication / consent in an HTTPs form to $\user$ \\
\quad \cmt{authentication and consent are often combined together}\\
\quad \textbf{If} $\user$ is authenticated \textbf{and} gives consent \textbf{then}\\
\quad \quad select $\codeB$ (by OAuth 2.0 standard)\\
\quad \quad send ($\msg{auth}, \codeB, \localstateB$) to $\mixer$\\
\quad \textbf{else} output failure\\
\onrecv{} ($\msg{token}, \clientidB, \clientsecB, \redirecturiB,$ \\ $\codeB'$) from $\mixer$:\\
\quad query $\clientsecB'$ and $\redirecturiB'$ using $\clientidB$ \\
\quad assert $\clientsecB' = \clientsecB$\\
\quad assert $\redirecturiB' = \redirecturiB$\\
\quad assert $\codeB' = \codeB$ \\
\quad select $\tokenB$ (by OAuth 2.0 standard)\\
\quad send ($\msg{token}, \tokenB$) to $\mixer$\\
\onrecv{} ($\msg{identity}, \tokenB'$) from $\mixer$:\\
\quad assert $\tokenB' = \tokenB$\\
\quad \pcreturn $\userid$
}
\end{boxedminipage}

    \caption{Login flow protocol for $\IDP$}
    \label{fig:protocolIdP}
\end{figure}

\begin{figure}
    \centering

\begin{boxedminipage}{\linewidth}

\begin{center}
    \scriptsize{$\PROT_{\RP}$ (Login flow)}
\end{center}

\procedure[mode=text, codesize=\footnotesize,linenumbering]{}{
\onrecv{} ($\msg{login}$) from $\user$:\\
\quad sample $\localstateRP \sample \ZZ$\\
\quad send ($\msg{auth},\clientidRP, \localstateRP, \redirecturiRP$) to $\mixer$\\
\onrecv{} ($\msg{auth}, \codeRP, \localstateRP'$) from $\mixer$:\\
\quad assert $\localstateRP' = \localstateRP$\\
\quad send ($\msg{token}, \clientidRP, \clientsecRP, \redirecturiRP, \codeRP$) to $\mixer$\\
\onrecv{} ($\msg{token}, \tokenRP$) from $\mixer$:\\
\quad send ($\msg{identity}, \tokenRP$) to $\mixer$\\
\quad wait to receive $\uid$
}
\end{boxedminipage}

    \caption{Login flow protocol for $\RP$}
    \label{fig:protocolRP}
\end{figure}

\section{\extendedprotocol{}}
\label{sec:extended}

We now specify the \mofn{m}{n} \extendedprotocol{}.
The user first selects $n$ IdPs to provide identities on the very first sign-on (this can be integrated in a user sign-up when deployed). After that, the user is supposed to select a combination of at least $m$ out of the $n$ IdPs in further SSO, where $m$ is the threshold of the scheme that is often smaller than $n$, e.g., in a \mofn{3}{4} \extendedprotocol{}, the user may select a combination of $3$ or $4$ IdPs after his or her first sign-on.

\begin{figure}
  \centering
  \includegraphics[width=\linewidth]{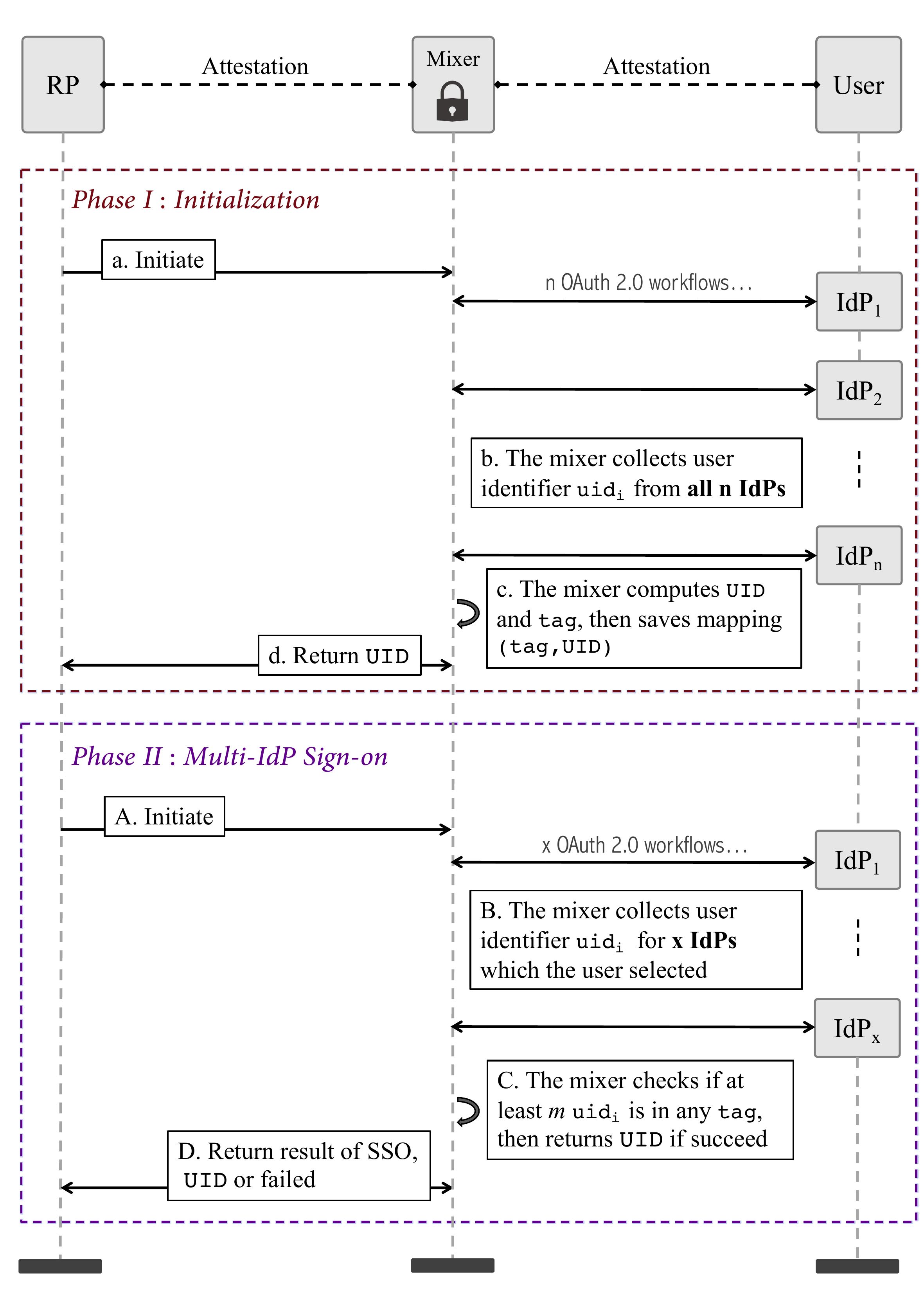}
  \caption{\extendedprotocol{} workflow for \name{}.}
  \label{fig:extended}
\end{figure}

All of the parties as well as their roles stay unchanged in comparison with the setting under a single IdP. The \extendedprotocol{} is also make up of the two registration steps and the login flow.
The registration steps, i.e., mixer registration and RP registration, shall be done before the \extendedprotocol{} login flow. 
For the mixer registration, the mixer $\mixer$ needs to register with all $n$ $\IDP$s, $\IDP_i, i = 1,2,...,n$, obtaining all $n$ client credentials, \clientidB{}${_{i}}$, \clientsecB{}${_{i}}, i = 1,2,...,n$.
The login flow of \mofn{m}{n} \extendedprotocol{} is presented in \autoref{fig:extended}. This login flow in \extendedprotocol{} contains two phases:

\begin{itemize}
    \item \emph{phase I.} The \emph{initialization} phase is executed when the user first sign-on $\RP$, and the mixer $\mixer$ interacts with all $n$ $\IDP$s and calculates a blinded identifier \uid{} for the user. 
    \item \emph{phase II.} The \emph{multi IdP sign-on} phase, which is executed for the user's rest signs-on, and the mixer $\mixer$ interacts with at least $m$ $\IDP$s the user selected. 
\end{itemize}

Each phase breaks into 4 similar parts.

\boldhead{Phase I: Initialization.}

\begin{enumerate}[label=\alph*.]
    \item $\RP$ initiates the flow. This is an authorization request. This part is identical to step \mysquare{1} in \autoref{fig:design-abstract} except for the URL parameter \idplist{} in the request, which contains all the names (or identifiers that helps $\mixer$ to identify) of the $n$ $\IDP$s.
    \item $\mixer$, user, and \textbf{n} $\IDP$s go through $n$ OAuth 2.0 workflows sequentially. For a single workflow involving $\IDP_i$, $\mixer$ collects the user's user identifier \userid{}$_i$ corresponding to $\IDP_i$. Readers can treat this part as a sequential execution of steps \mysquare{2} - \mysquare{7} in \autoref{fig:design-abstract}.
    $\mixer$'s \clientidB{}${_{i}}$, \clientsecB{}${_{i}}$, and \redirecturiB{} are used in each OAuth 2.0 workflow.
    \item $\mixer$ uses the $n$ user identifiers \userid{} to generate a \preuid{}:
    \[
    \preuid{} \coloneqq \mathsf{PRF}(\ktee{}, \userid{}_1\|\userid{}_2\cdots{}\|\userid{}_n\|\clientidRP{})
    \]
    Then, a per-user salt $\mathsf{s}$ is generates and saved to the encrypted disk as a mapping $(\preuid{}, \mathsf{s})$. 
    $\mixer$ also calculates a vector:
    \[
    \begin{split}
    \Vec{tag} \coloneqq [\mathsf{PRF}(\ktee{}, \userid{}_1),\mathsf{PRF}(\ktee{}, \userid{}_2),\cdots{},\\\mathsf{PRF}(\ktee{}, \userid{}_n)]
    \end{split}
    \]
    Then the \uid{} is generated as:
    \[
    \uid{} \coloneqq \mathsf{PRF}(\ktee{}, \userid{}_1\|\userid{}_2\cdots{}\|\userid{}_n\|\clientidRP{}\|\mathsf{s})
    \]
    The \uid{} is the one and only unique user identifier that will be passed to $\RP$ later to distinguish a user. 
    $\Vec{tag}$ containing pseudorandoms for all $n$ user identifiers \userid{}, and it will be used in phase II for retrieving the \uid{}.
    $\mixer$ seals $(\Vec{tag}, \uid{})$ in a table to the encrypted disk for future look-up.

    \item $\mixer$ returns \uid{} to $\RP$ as the result of this sign-on instance. This part is a combination of steps \mysquare{8} - \mysquare{12} in \autoref{fig:design-abstract}.
\end{enumerate}

In this phase, $\mixer$ joins one outer OAuth 2.0 workflow with $\RP$, and $n$ inner OAuth 2.0 workflow with the $n$ $\IDP$s and the user together.

\boldhead{Phase II: Multi-IdP Sign-on.}
\begin{enumerate}[label=\Alph*.]
    \item This is identical to step \emph{a} in phase I.
    \idplist{} should contain at least $m$ $\IDP$ names that the user selects to log in with.
    \item $\mixer$, the user and $x \quad  (m \leq x \leq n)$ $\IDP$s go through $x$ OAuth 2.0 workflows, collecting $x$ user identifiers \userid{}$_i$ with respect to $\IDP_i$.
    \item $\mixer$ computes $uid_i \coloneqq \mathsf{PRF}(\ktee{}, \userid{}_i)$ for all $x$ user identifiers \userid{} and compares them with all the $\Vec{tag}$s saved. For a certain $\Vec{tag}$, traverse all the elements in it, if at least $m$ out of the $x$ $uid_i$s are identical to those elements in this $\Vec{tag}$, we call this a successful match to a specific $(\Vec{tag}, \uid{})$.
    If succeed, $\mixer$ retrieves the corresponding \uid{} that bounded to it. Otherwise, phase II workflow fails.
    \item $\mixer$ returns the result of sign-on, \uid{} of the user if success, or failure.
\end{enumerate}

In this phase, $\mixer$ joins one outer OAuth 2.0 workflow with $\RP$, and (at least) $m$ inner OAuth 2.0 workflows with $\IDP$s and the user together.

In summary, the \extendedprotocol{} workflow is the \emph{extended} version of the basic SSO workflow and it displays an interesting application of \name{}. 
The basic SSO workflow under a single IdP can also be regarded as a special case, a \mofn{1}{1} \extendedprotocol{} in particular.
Besides, RPs should not force users to login with multiple IdPs. A more user-friendly approach is to allow users to choose flexibly whether to login with multiple IdPs. The \extendedprotocol{} solution is more suitable under the decentralized identity settings, where IdPs are less trusted, e.g., the Torus wallet~\cite{torus}.

\section{Security Definitions}

~\autoref{fig:prfattackgame} shows the PRF game and ~\autoref{fig:misoreductiongame} shows the reduction from the PRF game to distinguish $\mathsf{G}^{unlink}_{\name{}}$ and $\mathsf{G}^{unlink}_{\name{}'}$.

\begin{figure}
    \centering
    
\begin{pcimage}
\createprocedureblock{procb}{center,boxed}{}{}{linenumbering, width=\linewidth}
\procb{$\mathsf{G}^{PRF}(\secparam, \bdv{})$}{
\sk \sample \kgen(\secparam) \\
b \sample \bin \\
\textbf{if $b=0$, } f \gets \prf{}(\sk, \cdot) \\ 
\textbf{if $b=1$, $f \gets rand()$ \cmt{a random function}} \\
x_1, x_2, \cdots, x_i \gets \bdv \\
y_i \gets f(x_i), i = 1, 2, \cdots \\
b' \sample \bdv(y_1, y_2, \cdots, y_i) \\
\pcreturn b=b'
}
\end{pcimage}

    \caption{PRF security game.}
    \label{fig:prfattackgame}
\end{figure}

\begin{figure}
    \centering
\begin{pcimage}
\createprocedureblock{procb}{center,boxed}{}{}{linenumbering, width=\linewidth}
\procb{$\mathsf{G}^{PRF}(\secparam, \bdv{}, \adv{})$}{
\sk \sample \kgen(\secparam) \\
b \sample \bin \\
\textbf{if $b=0$, } f \gets \prf{}(\sk, \cdot) \\ 
\textbf{if $b=1$, $f \gets rand()$ \cmt{a random function}} \\
\clientidRP{}_m, \userid{}_n \gets \adv, m = 1, 2, \cdots, n = 1, 2, \cdots \\
x_{m, n} \coloneqq \userid{}_n\|\clientidRP{}_m\|s\\
x_{1,1}, x_{1,2}, \cdots, x_{m, n} \gets \bdv \\
y_{m, n} \gets f(x_{m, n}), m = 1, 2, \cdots, n = 1, 2, \cdots \\
\uid{}_{m, n} \coloneqq y_{m, n} \\
b' \sample \adv(\uid{}_{1,1}, \uid{}_{1,2}, \cdots, \uid{}_{m,n}) \\
\pcreturn b=b'
}
\end{pcimage}

    \caption{Reduction from the PRF game to distinguish $\mathsf{G}^{unlink}_{\name{}}$ and $\mathsf{G}^{unlink}_{\name{}'}$}
    \label{fig:misoreductiongame}
\end{figure}

\section{\name{} Prototype Demo}
\label{a:demo}

\begin{figure*}
  \centering
  \subfloat[RP front-page]{\includegraphics[width=0.4\textwidth]{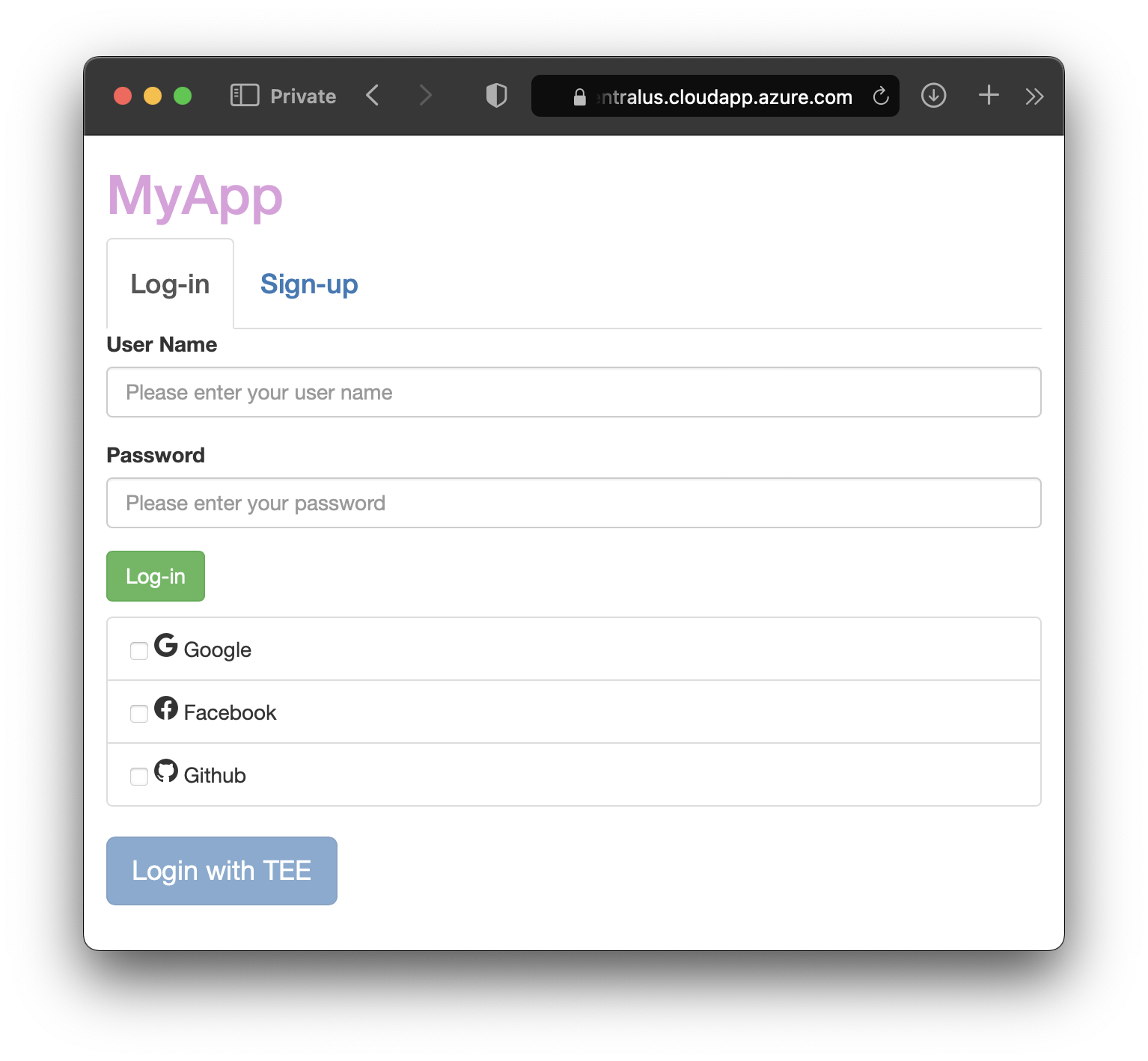}\label{fig:demo-frontpage}}
  \hspace{1cm}
  \subfloat[Sign-up (phase I)]{\includegraphics[width=0.4\textwidth]{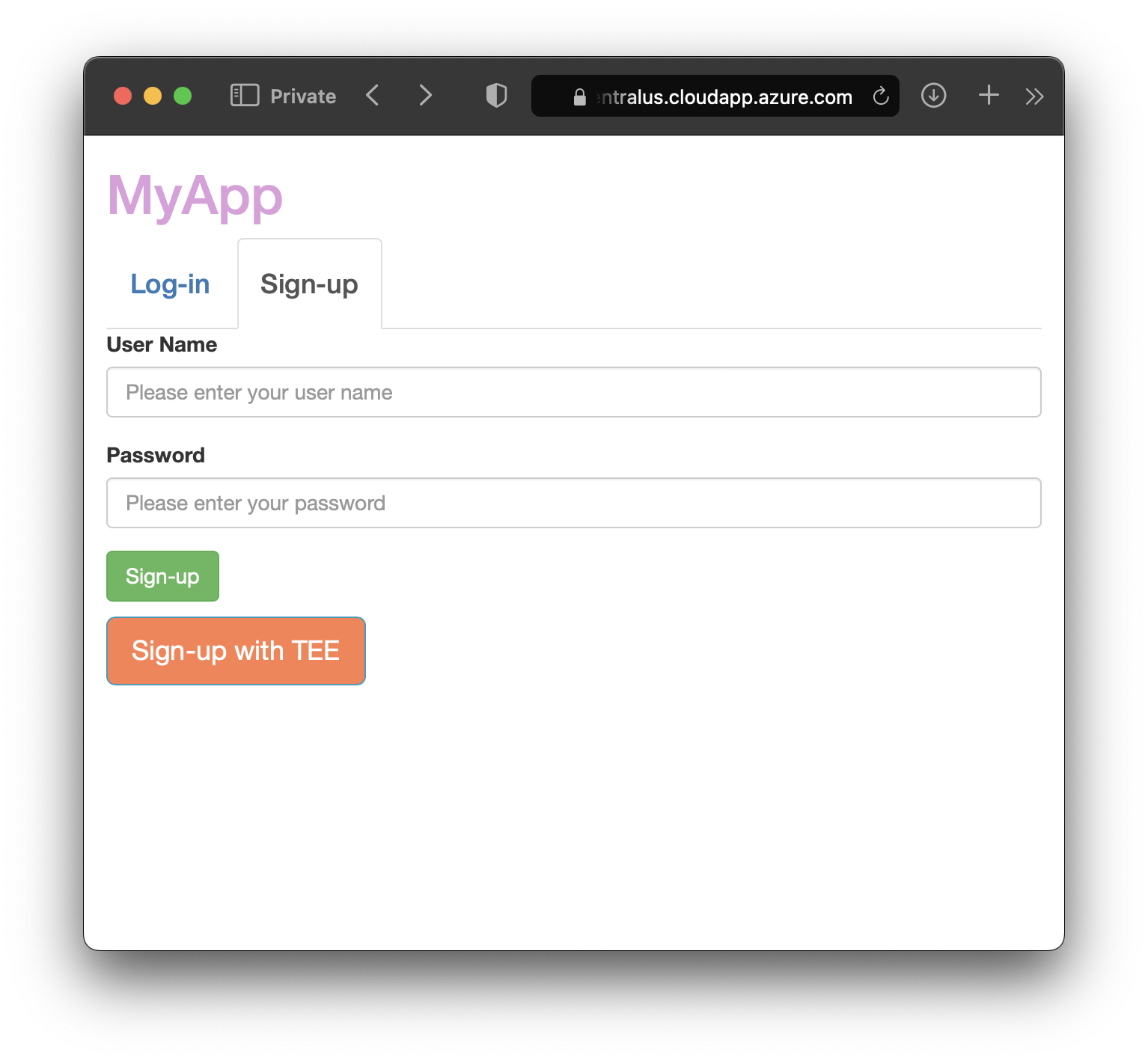}\label{fig:demo-phasei}} 
  \\
  \subfloat[User authentication with Google]{\includegraphics[width=0.32\textwidth]{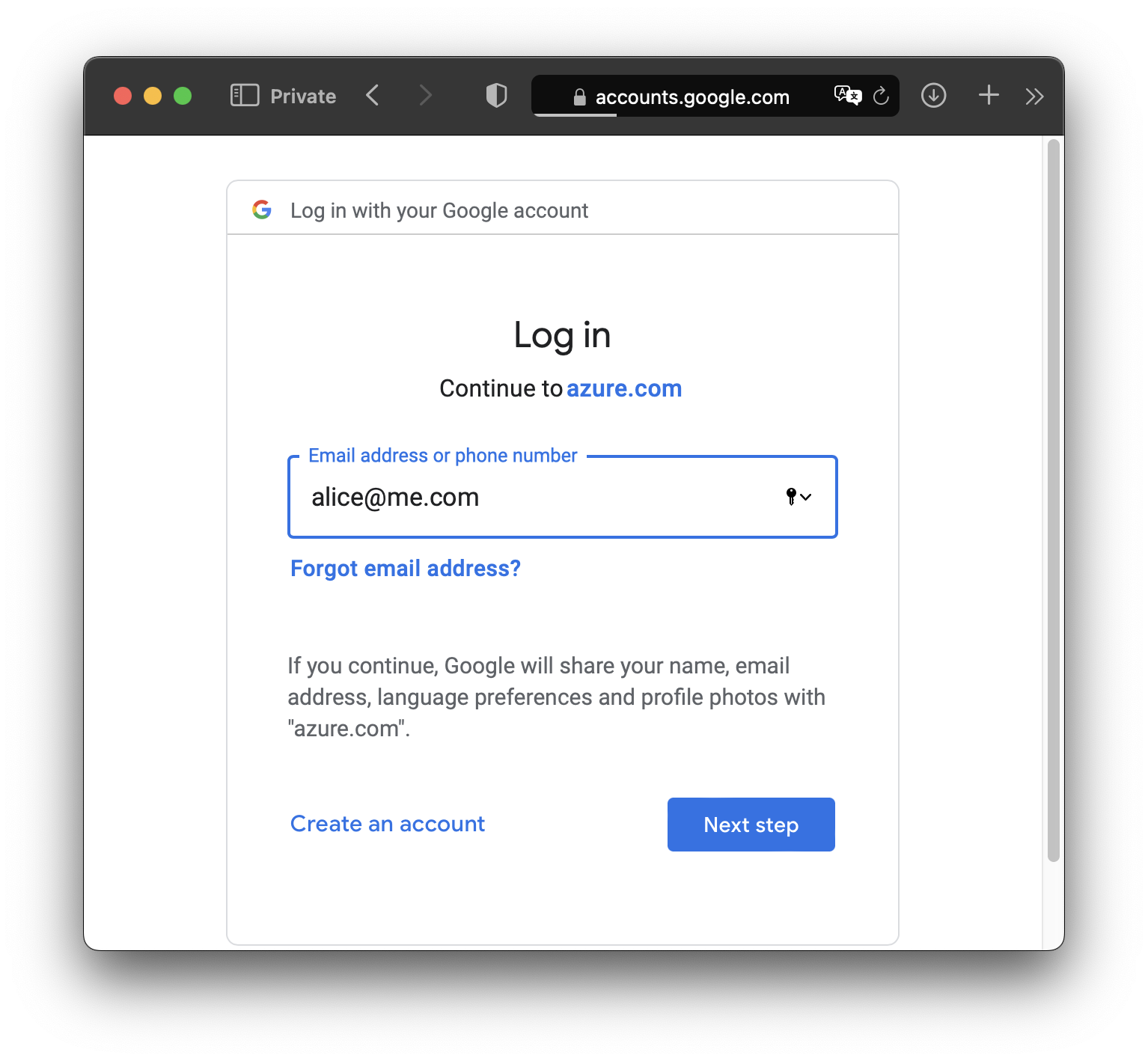}\label{fig:demo-google}} 
  \subfloat[User authentication with Facebook]{\includegraphics[width=0.32\textwidth]{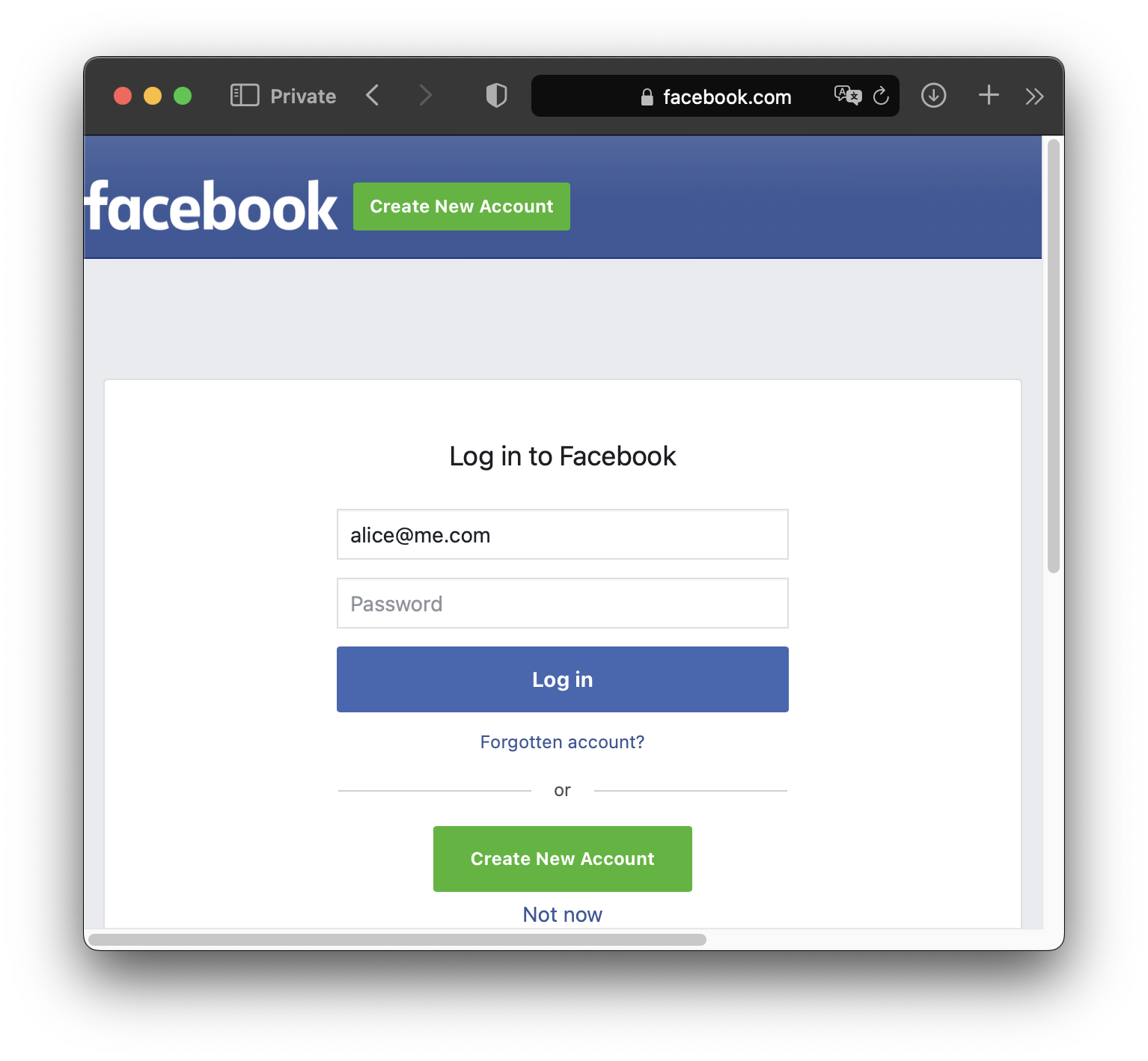}\label{fig:demo-facebook}} 
  \subfloat[User authentication with Github]{\includegraphics[width=0.32\textwidth]{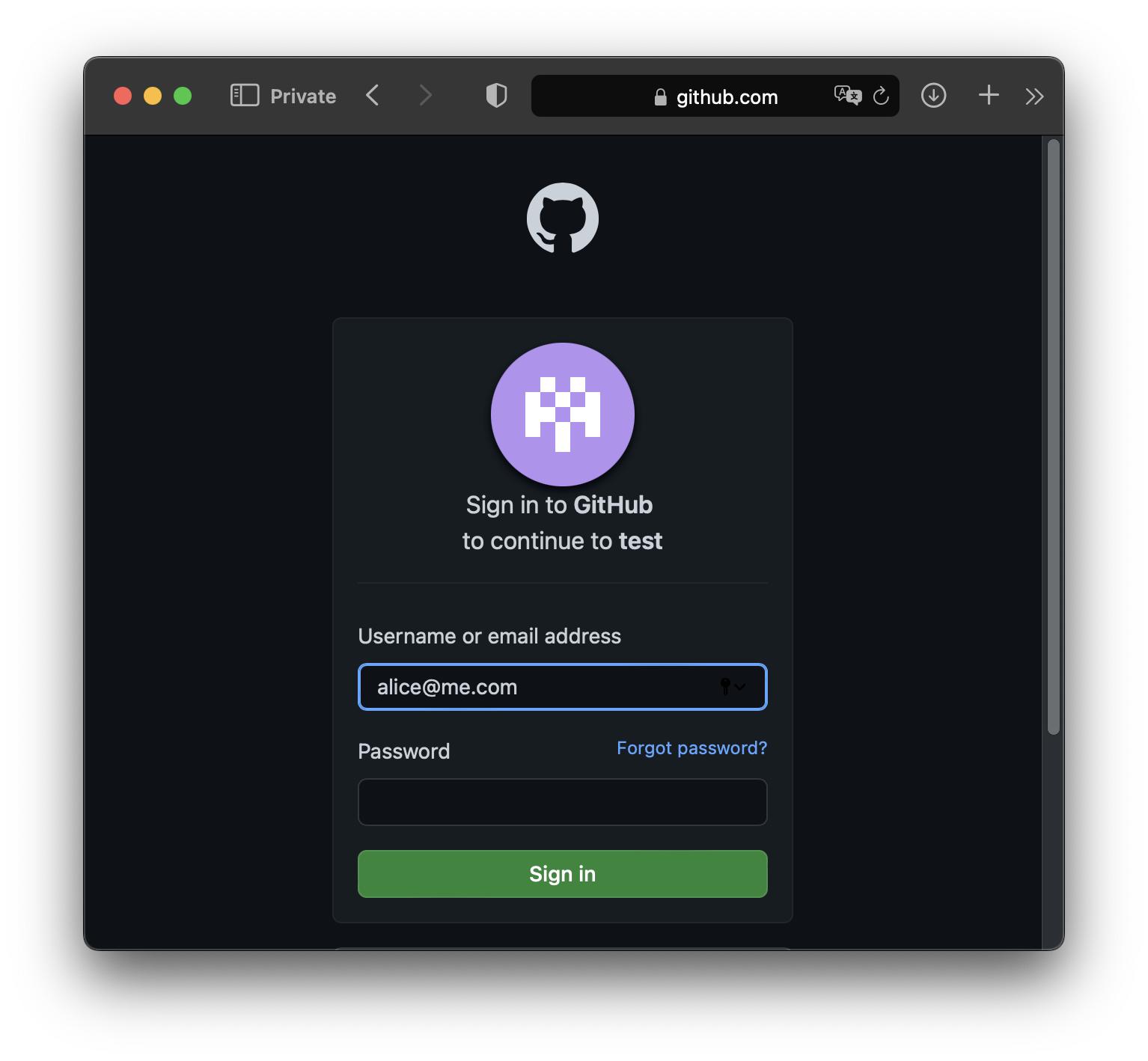}\label{fig:demo-github}} 
  \\
  \subfloat[Sign-up successfully]{\includegraphics[width=0.4\textwidth]{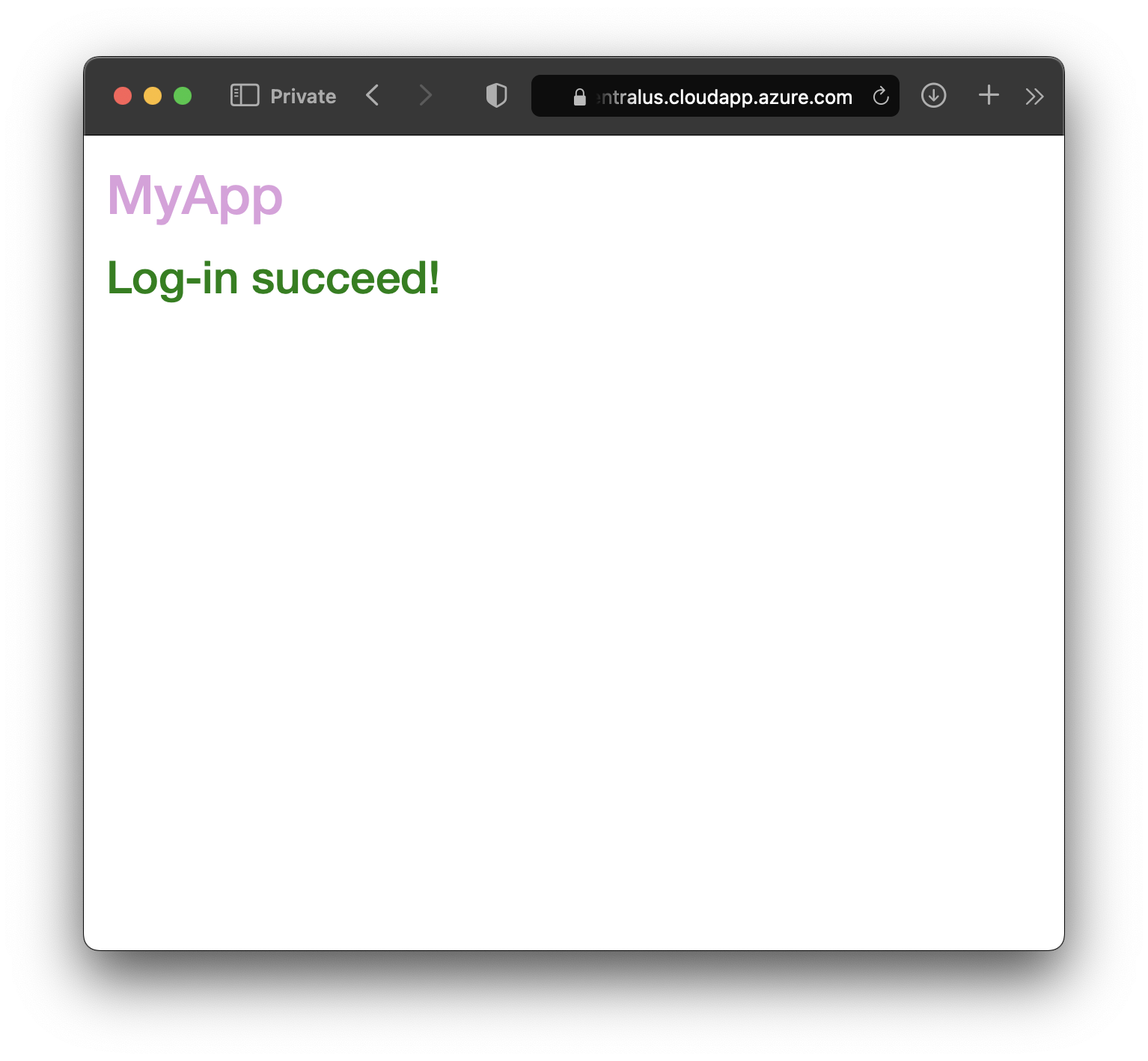}\label{fig:demo-success}} 
  \hspace{1cm}
  \subfloat[Login with 2 of 3 IdPs (phase II)]{\includegraphics[width=0.4\textwidth]{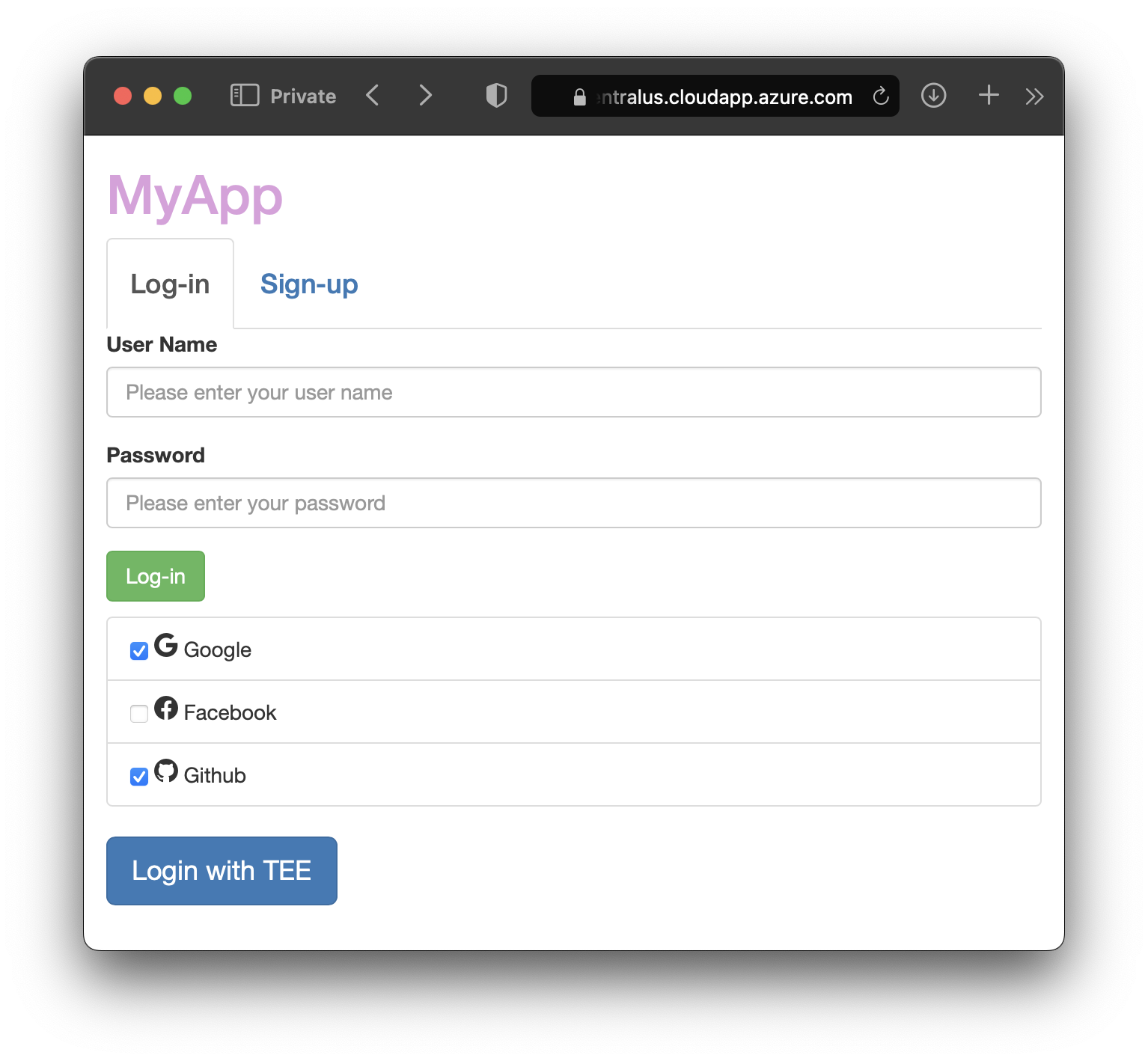}\label{fig:demo-phaseii}}
  \\
  \subfloat[Mixer server logs: RP exchanges the access token \tokenRP{} to the blinded user identifier \uid{}]{\includegraphics[width=0.4\textwidth]{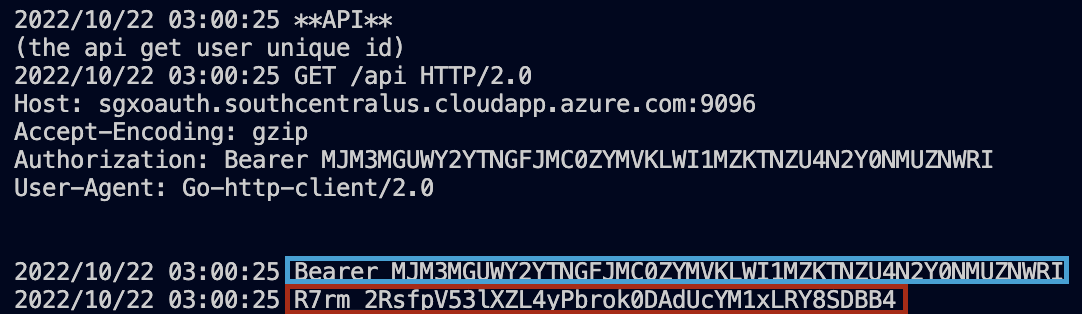}\label{fig:demo-logsTEE}}
  \hspace{1cm}
  \subfloat[RP server logs: \uid{}]{\includegraphics[width=0.4\textwidth]{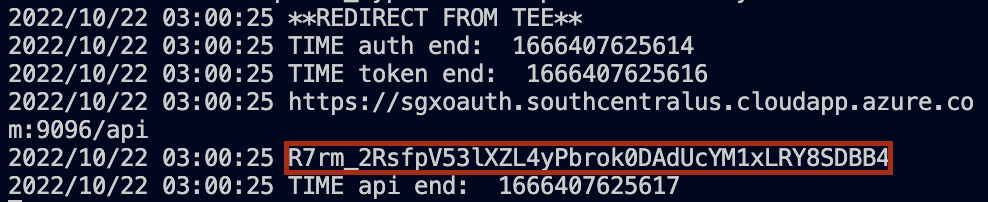}\label{fig:demo-logsRP}}
  \caption{Demo and logs of a \mofn{2}{3} Multi-IdP SSO login using \name{}.}
  \label{fig:demo}
\end{figure*}

We present the prototype demo for the \mofn{2}{3} \extendedprotocol{} with \name{} in \autoref{fig:demo}.
Suppose the user uses Apple Safari as the user-agent.
The mixer server host is started at \texttt{https://sgxoauth.southcentralus.cloudapp.\\azure.com:9096}.
We start an RP so-called \textcolor{plum}{MyApp}, its login front page is shown as \autoref{fig:demo-frontpage}.
For the \extendedprotocol{}, the user need to go through phase I for initialization (\autoref{fig:demo-phasei}). 
Instead of creating her account and password on the RP, the user clicks the orange \emph{Sign-up with TEE} button to initiate the SSO process with \name{}.
She then authenticates with all 3 IdPs in a row: Google (\autoref{fig:demo-google}), Facebook (\autoref{fig:demo-facebook}), and Github (\autoref{fig:demo-github}). 
Notice that those IdPs can fit \name{} without any modifications.
After these steps, the user will be logged into the RP successfully (\autoref{fig:demo-success}).
For the upcoming logins, she can select at least 2 IdPs on the front page as shown in \autoref{fig:demo-phaseii}, this is phase II of the \mofn{2}{3} \extendedprotocol{}. 
The mixer does not need to set up a front page, so it is hard to demonstrate its role in this demo.
We display the server logs instead.
\autoref{fig:demo-logsTEE} shows the server logs of the mixer, in which the RP calls the \texttt{/api} endpoint (resource endpoint \resourceendB) of the mixer to exchange \tokenRP{} to the blinded user identifier \uid{}.
In \autoref{fig:demo-logsRP}, the RP obtains the same \uid{} returned from the mixer.

\end{appendices}

\end{document}